\documentclass[journal]{IEEEtran}
\usepackage{amsmath,amssymb,breqn}

\usepackage{graphbox} 
\usepackage{cite}
\usepackage{comment}
\usepackage{authblk}
\graphicspath{ {imgs/} }
\providecommand*{\rot}[1]{\rotatebox[origin=c]{90}{#1}}

\ifCLASSINFOpdf
  
\else
  
\fi

\begin{document}

\title{Model-Based Iterative Reconstruction for One-Sided Ultrasonic Non-Destructive Evaluation}

%
%
\author[1]{Hani Almansouri}
\author[2]{Singanallur Venkatakrishnan}
\author[1]{Charles Bouman}
\author[2]{Hector Santos-Villalobos}
\affil[1]{Purdue University, 610 Purdue Mall, West Lafayette, IN 47907}
\affil[2]{Oak Ridge National Laboratory, One Bethel Valley Road, Oak Ridge, TN 37831}


\maketitle
\begin{abstract}

 One-sided ultrasonic non-destructive evaluation (UNDE) is 
extensively used to characterize structures that need to be inspected and maintained 
from defects and flaws that could affect the performance of power plants, such as nuclear power plants.
Most UNDE systems send acoustic pulses into the structure of interest, 
measure the received waveform and use an algorithm to reconstruct the quantity of interest. 
The most widely used algorithm in UNDE systems is the synthetic aperture focusing technique (SAFT) 
because it produces acceptable results in real time. 
A few regularized inversion techniques with linear models have been proposed which can improve on SAFT, 
but they tend to make simplifying assumptions that
do not address how to obtain reconstructions from large real data sets. 
In this paper, we propose a model-based iterative reconstruction (MBIR) algorithm designed for scanning UNDE systems. 
To further reduce some of the artifacts in the results, we enhance the forward model to account for the transmitted beam profile, 
the occurrence of direct arrival signals, and the correlation between scans from adjacent regions. 
Next, we combine the forward model with a spatially variant prior model to account for the attenuation
of deeper regions. 
We also present an algorithm to jointly reconstruct measurements from large data 
sets. 
Finally, using simulated and extensive experimental data, we show MBIR results and demonstrate how 
we can improve over SAFT as well as existing regularized inversion techniques. 

\footnote{This manuscript has been authored by UT-Battelle, LLC under Contract No. DE-AC05-00OR22725 with the U.S. Department of Energy.  The United States Government retains and the publisher, by accepting the article for publication, acknowledges that the United States Government retains a non-exclusive, paid-up, irrevocable, world-wide license to publish or reproduce the published form of this manuscript, or allow others to do so, for United States Government purposes.  The Department of Energy will provide public access to these results of federally sponsored research in accordance with the DOE Public Access Plan (http://energy.gov/downloads/doe-public-access-plan).}
\end{abstract}


\IEEEpeerreviewmaketitle

\section{Introduction}
\begin{figure}
	\centering
	\includegraphics[width=0.3\textwidth]{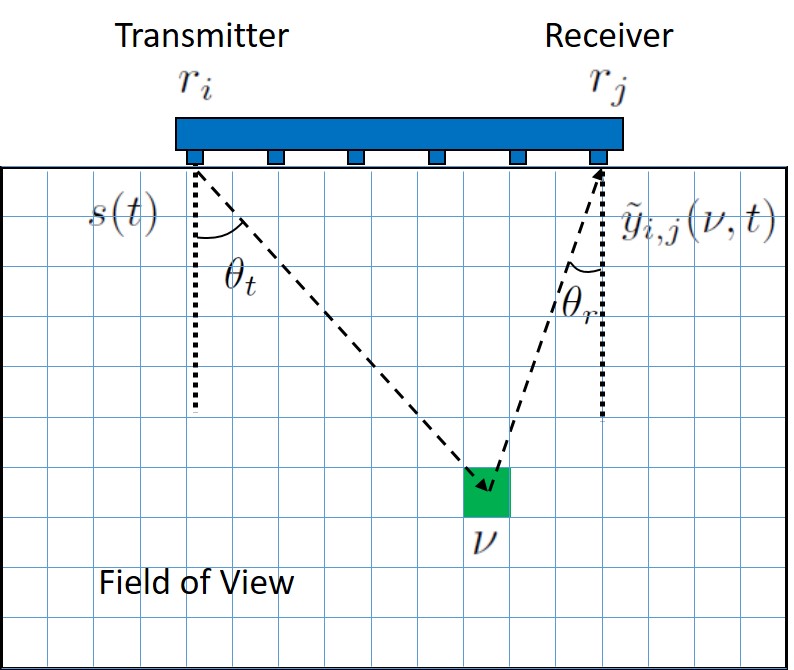}
	\caption{An illustration of a typical one-sided UNDE problem where $s(t)$ is the transmitted signal, $\nu$ is a point in the field-of-view, $y_{i,j}(\nu,t)$ is the received signal reflected from $\nu$, $\theta_t$ is the angle between $r_i$ and $\nu$, and $\theta_r$ is the angle between $r_j$ and $\nu$.}
	\label{fig:illu}
\end{figure}

\IEEEPARstart{O}{ne-sided} ultrasonic non-destructive evaluation (UNDE) is
widely used in many applications to characterize and detect flaws in materials, such as 
concrete structures in nuclear power plants (NPP), because of its low cost, high penetration, portability, and safety compared with other NDE methods \cite{Ramuhalli_2012,berndt_2001,Zemanek1970}.
A typical one-sided UNDE system consists of a sensor that transmits sound waves into the structures of interest and an array of receivers that measures the reflected signals (see Fig. \ref{fig:illu}). 
Such a set up is scanned across a large surface in a rectangular grid pattern and the reflected signals from each position are processed to reconstruct the underlying structure. 
The ability to easily probe structures that can only be accessed from a single side combined along with the ability of  ultrasound signals to penetrate deep into structures make one-sided UNDE a powerful tool for the analysis of structures across a variety of applications \cite{hoegh_khazanovich_2015,stepinski2007implementation}.

\begin{figure*}
	\begin{center}\footnotesize
		\begin{tabular}{@{}c@{}c@{}c@{}}
			\includegraphics[align = c,width=0.3\textwidth]{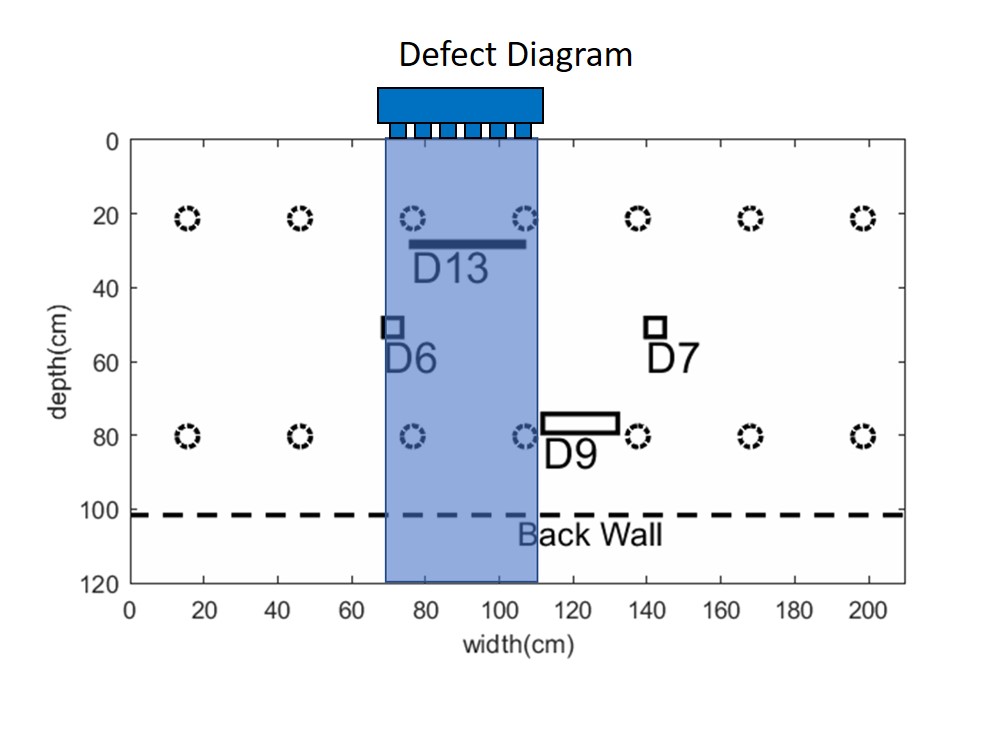} &
			\includegraphics[align = c,width=0.3\textwidth]{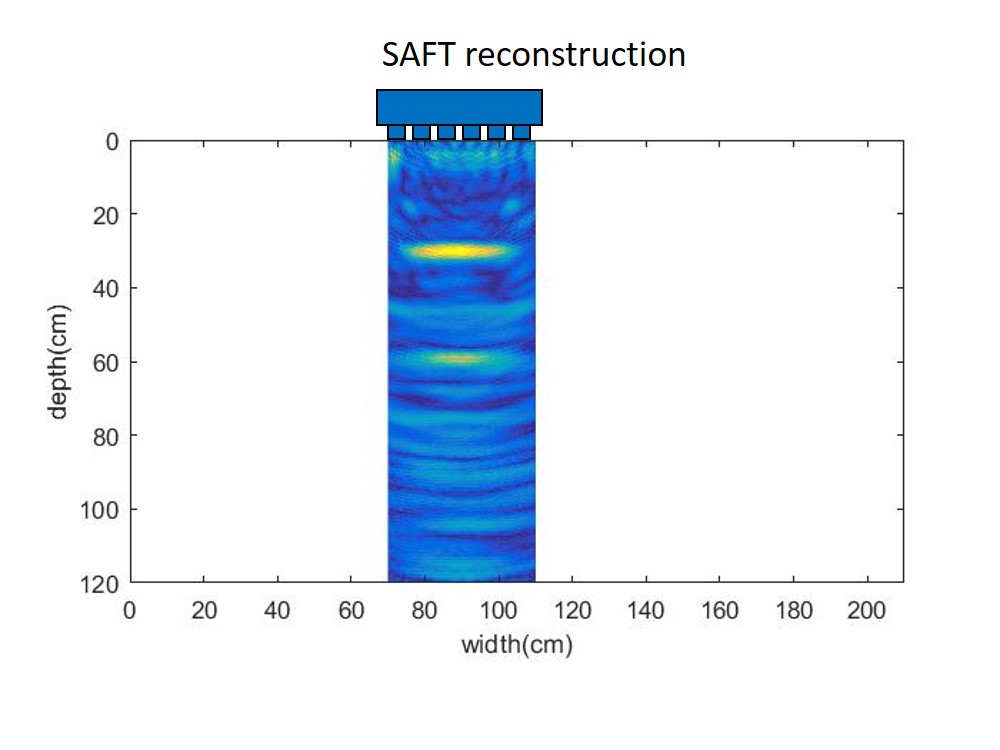}&
			\includegraphics[align = c,width=0.3\textwidth]{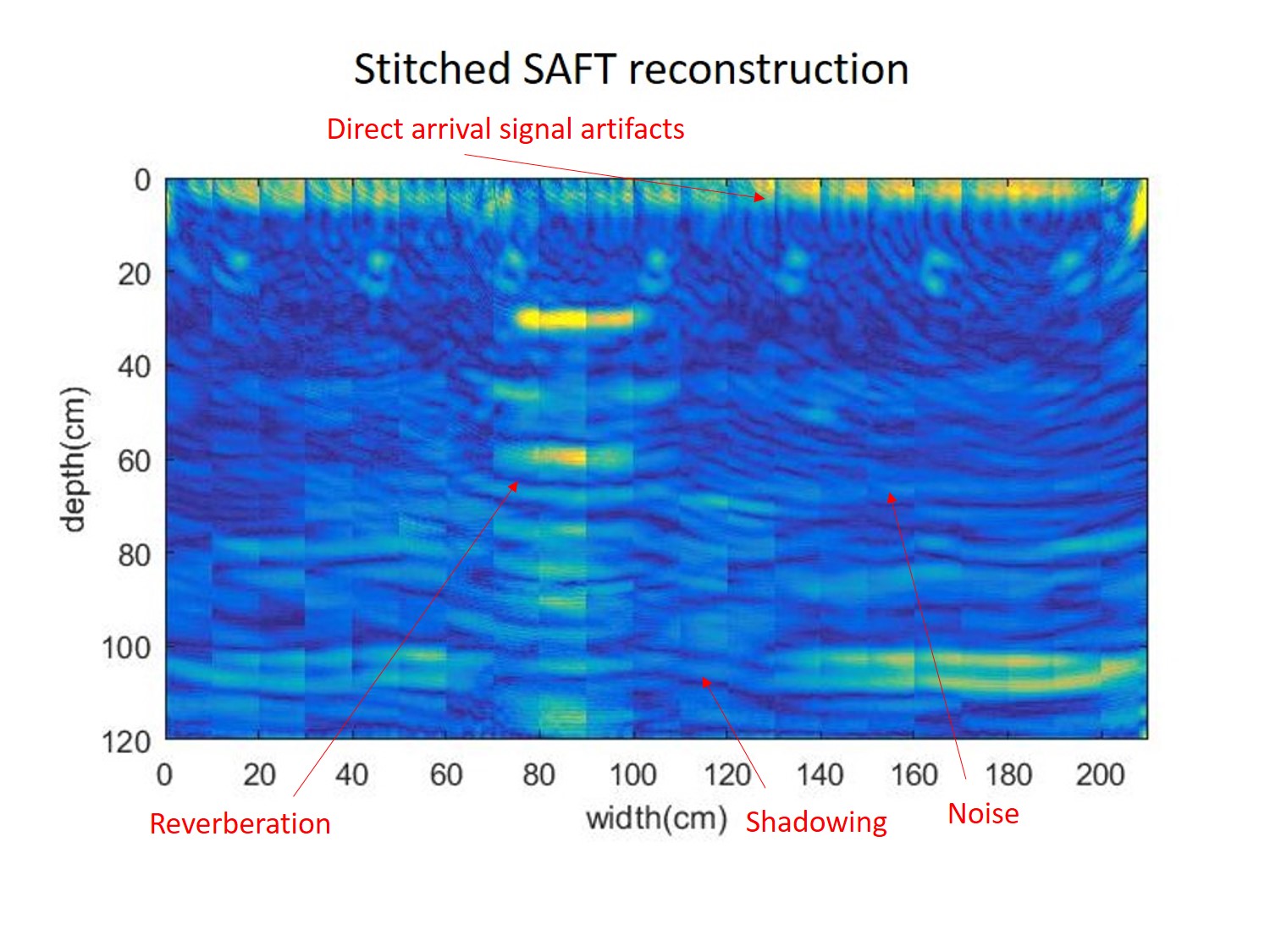}
			\tabularnewline
			(a) & (b) & (c)
			\tabularnewline
		\end{tabular}
	\end{center}
	\vspace{-0.15in}
	\caption{\label{fig:illu_SAFT}
		\scriptsize
		{Example of a SAFT reconstruction from real data of a concrete structure. (a) shows the defect diagram containing steel rebars (dotted circles), defects (marked D\#), and the back wall (dotted line). (b) shows SAFT reconstruction for a single scan of the large field-of-view in (a). (c) shows the SAFT reconstruction for the entire field-of-view after stitching the results from each individual scan.}}
\end{figure*}

Reconstruction of structures from one-sided UNDE systems are challenging because of the complex interaction of ultrasound waves with matter, the geometry of the experimental set-up, the trade-off between resolution and penetration, and the potentially low signal-to-noise ratio of the received signals \cite{li_hayward_2011,Haldorsen2006}. 
The most widely used reconstruction method for UNDE is the synthetic aperture focusing technique
(SAFT) \cite{shao2011,Engle2014,dobie2013,beniwal_ganguli_2015,hoegh_khazanovich_2015,schickert2003ultrasonic}.
SAFT uses a delay-and-sum (DAS) approach to reconstruct ultrasound images.
Typically, the measurements from each scan are processed independently
and stitched together thereby not accounting for the the effects of adjacent regions. 
Fig. \ref{fig:illu_SAFT} shows an example of a SAFT reconstruction from real data.
Notice that SAFT reconstructions tend to have significant artifacts due to the fact that SAFT assumes a simple propagation model and does not account for a variety of effects such as noise and image statistics, direct arrival signal artifacts, reverberation, and shadowing \cite{beniwal_ganguli_2015,schickert2003ultrasonic}. 
Furthermore, since each SAFT scan is independently processed, there are explicit
``stitching'' artifacts present in the reconstruction of a large cross-section.
In summary, while SAFT is computationally inexpensive to implement, it can result in significant artifacts in the one-sided UNDE reconstructions.

In order to overcome some of the short-comings of the SAFT method,
regularized iterative reconstruction methods that use linear models
(due to their low computational complexity)
have recently been proposed for various ultrasound inverse problems.
These methods formulate the reconstruction as minimizing a cost-function that balances a data fidelity term with a regularization applied to the image/volume to be reconstructed.
The data fidelity term encodes a physics based model to reduce the error
between the measurements and the projected reconstruction while the
regularizer forces certain constraints on the reconstruction itself.
For the data fidelity term, regularized iterative techniques for one-sided UNDE, such as \cite{ozkan2018inverse,wu2015model}, use a simple linear model that models the propagation of the ultrasonic wave to reconstruct the reflectance B-mode images.
A technique that uses the same forward model, but shows 2D images for a fixed depth (c-mode), is shown in \cite{tuysuzoglu2012sparsity}.
The forward model in \cite{tuysuzoglu2012sparsity} has been upgraded to account for the beam profile as in \cite{guarneri2015sparse} which can help in reducing some artifacts.
However, this forward model does not account for direct arrival signals caused by coupling the ultrasonic device to the surface of the structure which might cause artifacts and interference with reflections.
Furthermore,  the reconstruction algorithm of \cite{guarneri2015sparse} is not designed to exploit correlations between adjacent scans for systems with large field-of-view.

In \cite{szasz2016beamforming,guarneri2015sparse,wu2015model,shieh2012resolution}, the authors used a simple regularization terms, such as $l_1$ or $l_2$.
This regularization is suitable for imaging point scatters or sparse regions.
However, for more complex medium where edge preservation is needed, other techniques use a more sophisticated regularization, such as total variation, where they showed great enhancement over SAFT \cite{ozkan2018inverse,tuysuzoglu2012sparsity}.
The method in \cite{ozkan2018inverse} uses total variation with variety of a regularization terms that are depth dependent to resolve the attenuation and blurring for deeper reflections.
However, the depth-dependent regularization is linear with depth which might not be the best modeling for the depth attenuation. 
Therefore, while regularized inversion methods that use a linear forward model have shown promise in certain applications,
they do not deal with the direct arrival signal artifacts in a principled manner,
they have not been designed to jointly handle large data sets that 
require multiple scanning for one-sided UNDE systems, 
and they do not fully account for the depth-dependent blurring
that can occur by the use of certain regularizers.

%
%
%

In this paper, we propose an ultrasonic model-based iterative reconstruction (MBIR) algorithm designed specifically for one-sided UNDE systems of large structures.
We resolve the issues discussed above by enhancing the forward and prior models used in the current regularized iterative techniques.
The enhancement to the forward model include a direct arrival signal model with varying acoustic speed and an anisotropic modeling of the transmitted signal propagation to reduce some of the artifacts in the reconstruction.
Also, we repopulate the system matrix of the forward model to generate a larger system matrix for larger field of views to share more information about adjacent scans which can help in reducing noise and artifacts and enhancing the reconstruction.
Furthermore, the prior model is enhanced by increasing and conveniently controlling the regularization for deeper regions to reduce the attenuation to these regions.
In previous work, we have demonstrated the performance of MBIR compared with SAFT using different combinations of these enhancements \cite{almansouri2017progress,almansouri2018anisotropic,almansouri2018Ultrasonic}.
We introduce four major contributions in this paper:

{\addtolength{\leftskip}{5mm}	

	\noindent
	1) A physics-based linear forward model that models the direct arrival signal with varying acoustic speed, absorption attenuation, and anisotropic propagation;\\
	2) A non-linear spatially-variant regularization to enhance the reconstruction for deeper regions;\\
	3) A systematic way using joint-MAP stitching and 2.5D MBIR to reconstruct the volume from all the measured data simultaneously rather than individual reconstruction;\\
	4) Qualitative and quantitative results from simulated and extensive experimental data.
}
\noindent

The paper is organized as follows. 
In section \ref{forward} we cover the design for the forward model of the ultrasonic MBIR for one-sided NDE applications. 
In section \ref{prior_model} we cover the prior model used for MBIR. 
In section \ref{Opt} we cover the optimization of the MAP cost function using the ICD method. 
In section \ref{results} we cover simulated and experimental results from MBIR and other techniques. 
In section \ref{conc} we cover the conclusion. 

\section{Forward Model of One-Sided UNDE}
\label{forward}
The reconstruction in an MBIR setting is given by the following minimization problem,
\begin{eqnarray*}
	x_{MAP} = \underset{(x)}{\text{arg\,min}} \left\lbrace -\log p(y|x) -\log p(x)\right\rbrace,
\end{eqnarray*}
where $x$ is the image to be reconstructed, $y$ is the measured data, $x_{MAP}$ is the reconstructed image, $p(y|x)$ is the forward model and the probability distribution of y given x, $p(x)$ is the prior model and the probability distribution of x. 
The forward model is designed in the following way.
We will consider a one-sided UNDE for a concrete structure where the transducers are coupled to the surface as shown in Fig. \ref{fig:illu}.
We will consider a pressure signal (Pascal) transmitted from transducer 
$i$ located at position $r_i \in \mathbb{R}^3$, reflected by a point 
located at $\nu \in \mathbb{R}^3$, and received by transducer $j$ 
located at $r_j \in \mathbb{R}^3$.
We assume the Fourier transform of the temporal impulse response of a system sending a signal from $r_i$ and receiving from $\nu$ to be 
$$
G( r_i , \nu , f) = \lambda e^{ - ( \alpha (f) + j \beta(f) ) \| \nu -r_i \| } 
$$
where $ \lambda$ is a transmittance coefficient, 
$$
\alpha (f) = \alpha_0 |f| \quad \quad (\text{m}^{-1})
$$
is the rate of attenuation,
$$
\beta(f) = \frac{ 2 \pi f }{ c } \quad \quad(\text{m}^{-1}) 
$$
is the phase delay due to propagation through the specimen, and $c$ is the speed of sound \cite{Kak1978,Norton1981,wiskin_1997,huttunen_malinen_kaipio_white_hynynen_2005,voigt_2005,treeby_cox_2009,treeby_tumen_cox_2011}.
Similarly, we assume the Fourier transform of the impulse response of a system sending a signal from $\nu$ and receiving from $r_j$ to be 
$$
G( \nu, r_j , f) =  \lambda e^{ - ( \alpha (f) + j \beta(f) ) \|  r_j - \nu \|} \ .
$$ 
Assuming $s(t)$ (Pascal) is the input to the system and $\tilde{x}(\nu)$ ($\text{m}^{-3}$) is 
the reflectivity coefficient for $\nu$, then the output $\tilde{Y}_{i,j} ( \nu, f ) $ ($\text{Pascal} \cdot \text{m}^{-3} \cdot \text{Hz}^{-1}$) at 
the receiver due to $\nu$ is
\begin{eqnarray*}
	\label{eq::Y}
	\tilde{Y}_{i,j} ( \nu, f ) 
	&=& -S(f) G( r_i , \nu , f) \tilde{x}(\nu) G( \nu, r_j , f) \\
	&=&  -\lambda^2 \tilde{x}(\nu) S(f) e^ {- ( \alpha_0 c |f| + j 2 \pi f ) \tau_{i,j}(\nu)},
\end{eqnarray*}
where
\begin{eqnarray*}
	\tau_{i,j}(\nu) &=& \frac{ \| \nu -r_i \| + \| \nu -r_j \| }{ c } \quad \quad (\text{s}).
\end{eqnarray*}
By defining 
\begin{eqnarray}
	\tilde{h}(\tau_{i,j}(\nu), t) = {\cal F}^{-1} \left\{ -\lambda^2 S(f) e^{ - \alpha_0 c |f| \tau_{i,j}(\nu) }\right\}, \label{htilde}
\end{eqnarray}
where ${\cal F}^{-1}$ is the inverse Fourier transform, the time domain output signal, $\tilde{y}_{i,j}( \nu, t)$ ($\text{Pascal} \cdot \text{m}^{-3}$), is given by
\begin{eqnarray*}
	\tilde{y}_{i,j} ( \nu, t) =  \, \tilde{h}(\tau_{i,j}(\nu), t - \tau_{i,j}(\nu) ) \, \tilde{x}(\nu).
\end{eqnarray*}
Note that $\tilde{h}(\tau_{i,j}(\nu), t)$ is a function of $\tau_{i,j}$ and 
$t$, i.e. not directly a function of $\nu$.
This is a very useful property that can reduce the computational 
cost of evaluating $\tilde{h}$.
In many cases, $\tilde{h}(\tau,t)$ for any $\tau$ is close to zero after a certain time $t_0$.
In this case, it is very helpful to modify the previous equation to
\begin{eqnarray*}
		\tilde{y}_{i,j} ( \nu, t) =  \, h(\tau_{i,j}(\nu), t - \tau_{i,j}(\nu) ) \, \tilde{x}(\nu). \\
\end{eqnarray*}
where 
\begin{eqnarray*}
	 h(\tau, t) &=& \tilde{h}(\tau, t ) \ \text{rect}\left( \frac{t}{t_0}-\frac{1}{2}\right), \\ \\
 \text{rect}(x) &=& 1 \text{ for }  |x| < \frac{1}{2} \text{ and 0 for } |x| \geq \frac{1}{2},
\end{eqnarray*}
and $t_0$ is a constant where we assume $h(\tau,t)$ is equal to zero for $t>t_0$.
Applying the rect function is very helpful in increasing the sparsity of the system matrix which leads to a dramatic decrease in memory and processing time.
To get the overall output $\tilde{y}_{i,j} (t) $ (Pascal) from all points in $\mathbb{R}^3$, we 
need to integrate over all $\nu$:
\begin{eqnarray}
\tilde{y}_{i,j} (t) 
&=& \int_{\mathbb{R}^3} \tilde{y}_{i,j} ( \nu, t) d\nu \\
&=& \int_{\mathbb{R}^3} \tilde{A}_{i,j}(\tau_{i,j}(\nu),t) \tilde{x}(\nu) d\nu \ , \label{forwardmodel}
\end{eqnarray}
where 
\begin{eqnarray}
\tilde{A}_{i,j}(\tau_{i,j}(\nu),t) =  h(\tau_{i,j}(\nu), t - \tau_{i,j}(\nu) ) . \label{Acont}
\end{eqnarray}
For simplicity, the set of all transducer pairs, $\{i,j\}$, is 
mapped to the ordered set $\{1,...,K\}$ , where $K$ is the total 
number of transducer pairs.
Hence, Eq. \ref{forwardmodel} becomes
\begin{eqnarray}
\tilde{y}_k (t) = \int_{\mathbb{R}^3} \tilde{A}_k(\tau_k(\nu),t) \tilde{x}(\nu) d\nu \ . \label{forwardmodel2}
\end{eqnarray}
Finally, we assume the noise associated with the measurements to be i.i.d. Gaussian.

\subsection{Direct Arrival Signal Artifacts}
\label{shift_error}
When the ultrasonic device is attached or coupled to the surface of the concrete, a direct arrival signal is generated along with the transmitted signal.
This direct arrival signal produces artifacts on the reconstructed image in regions closer to the transducer and it might interfere with some of the reflected signals (see Fig. \ref{fig:illu_SAFT}).
Eq. \ref{forwardmodel2} models the output from the reflection of all points. 
However, the equation does not account for the direct arrival signal.
Locating and deleting the direct arrival signal from the received signal 
eliminates the artifacts, but might lead to deleting reflection 
signals for closer objects.
We propose a modification to the forward model that models the 
direct arrival signal and attenuates the artifact while 
preserving information from reflected signals.
The modification adds the following term to the forward model in Eq. \ref{forwardmodel2} that corresponds to the direct arrival signal,
\begin{eqnarray}
\tilde{y}_k (t) =  \int_{\mathbb{R}^3} \tilde{A}_k(\tau_k(\nu),t) \tilde{x}(\nu) d\nu + \tilde{d}_k(t) \ g_k, \label{das}	
\end{eqnarray}
where $\tilde{d}_k(t)$ is an additional term used to model the direct arrival signal signal given by
\begin{eqnarray*}
\tilde{d}_k(t) &=& -\tilde{A}_k(\tau_k,t), \\
\tau_k &=& \frac{\|r_i-r_j \|}{c},
\end{eqnarray*}
and $g_k$ is an unknown scaling coefficient for the direct arrival signal. 

The above model works efficiently when the acoustic speed is constant. 
For a non-homogeneous material, such as concrete, the acoustic 
speed is not constant.
This change in acoustic speed changes the location of the direct arrival signal and causes a mismatch with MBIR's direct arrival signal modeling.
We can estimate the shift error by searching for the delay that produces the maximum autocorrelation of the direct arrival signal,
\begin{eqnarray*}
	\hat{l} &=&  \underset{-\tilde{\tau} \leq l \leq \tilde{\tau}} {\text{arg\,min}} \left\{\int  \tilde{y}_k(t) \tilde{d}_k(t-l) dt\right\}\\
	\tilde{d}_k(t) &\leftarrow& \tilde{d}_k(t-\hat{l}),
\end{eqnarray*}
where $\tilde{\tau}$ is chosen to be small, e.g. 3 sampling periods, to insure the shift is within the integral boundaries and to avoid interfering with later reflections.
This estimate finds the shift error with the assumption that reflections do not interfere with the direct arrival signal.
Therefore, for homogeneous medium, our approach is able to reduce direct arrival signal artifacts and detect reflections close to the transducers.
However, for non-homogonous medium, our approach is able to reduce direct arrival signal artifacts that do not interfere with reflections.

\subsection{Anisotropic Propagation}
\begin{figure}
	\centering
	\includegraphics[width=0.5\textwidth]{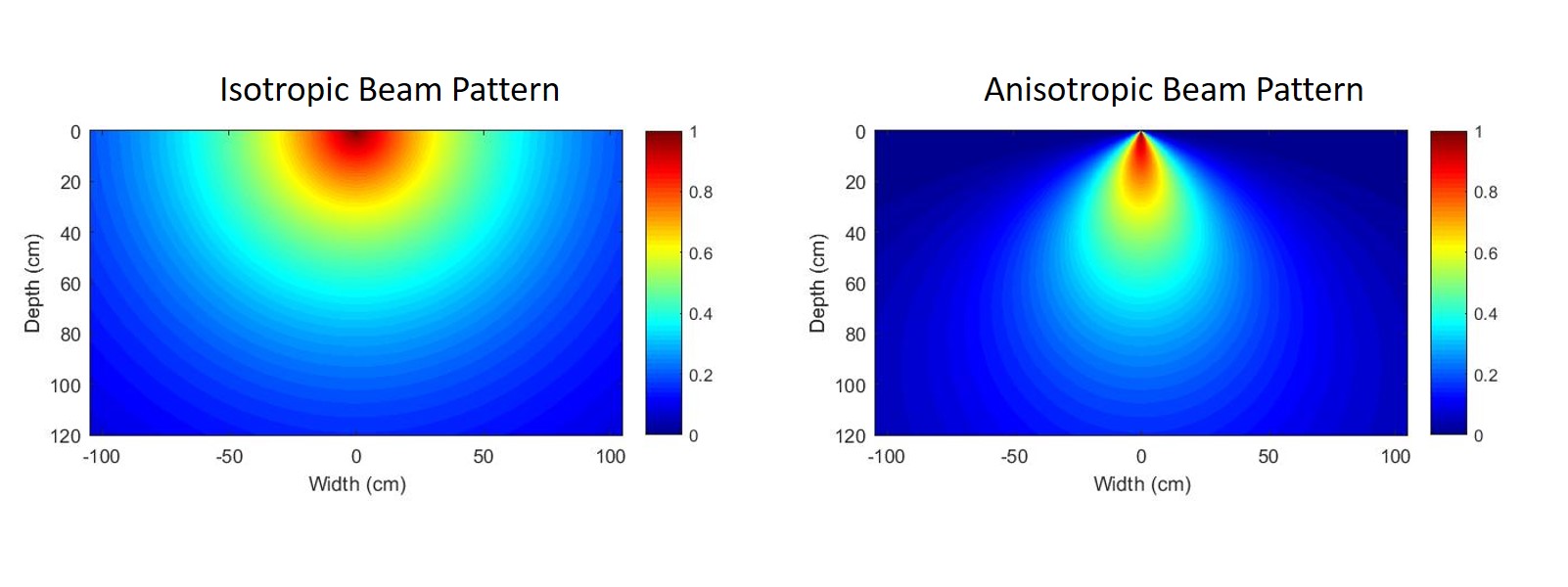}
	\caption{Beam pattern model for an ultrasound transducer placed at (0,0) for isotropic propagation (left) and anisotropic propagation (right). Left image shows equal propagation in all direction. Right image shows more attenuation as the angle between the transmitter and the pixel increases.}
	\label{aniso_propagation} 
\end{figure}

Many models used in UNDE assume that the profile of the transmitted beam is isotropic\cite{tuysuzoglu2012sparsity,li2011ultrasound}.  
However, this assumption is not valid for many systems and it can produce artifacts. 
While it would be ideal to know the precise profile especially of the transmitted beam, in systems that we deal with, this is not known. 
Therefore, we adopt a similar apodization function as in \cite{hoegh_khazanovich_2015} for the anisotropic model.
However, the apodization function used in \cite{hoegh_khazanovich_2015} has a slow attenuating window.
In our application, a faster attenuating window is needed.
We use an anisotropic beam pattern model as shown in Fig. \ref{aniso_propagation}.
We define a function, $\phi_k(\nu)$, that has a value ranging from $0$ to $1$. 
This function depends on the angles from the transmitter to $\nu$ and from $\nu$ to the receiver.  
$\phi_k(\nu)$ is monotonically decreasing with respect to those two angles. 
$\phi_k(\nu)$ can act as an attenuating window, such as cosine or Gaussian windows, to the output. 
$\phi_k(\nu)$ is added to Eq. \ref{Acont} as follows:
\begin{eqnarray}
\tilde{A}_k(\tau_k(\nu),t) = h(\tau_k(\nu), t - \tau_k(\nu) ) \phi_k(\nu)  
\end{eqnarray}
Note that the beam pattern is assumed to be reciprocal, i.e. the receiver will also have the same beam pattern.
In this paper, we chose $\phi_k(\nu)$ to be 
$$
\phi_k(\nu) = \cos^2(\theta_t(\nu)) \cos^2(\theta_r(\nu)) \ \ ,
$$
where $\theta_t$ is the angle between the transmitter and $\nu$ and $\theta_r$ is the angle between the receiver and $\nu$ shown in Fig. \ref{fig:illu}.

Finally, the discretized version of the forward model can be used in the MAP estimate as shown below,
\begin{eqnarray*}
	\begin{split}
		-\log p(y|x) =  \frac{1}{2\sigma^2} \left\|  y-Ax-Dg\right\|^2 + \text{constant},
	\end{split}
\end{eqnarray*}
where $y \in \mathbb{R}^{MK\times 1}$ is the measurement, $A \in \mathbb{R}^{MK\times N}$ is the forward model (system matrix), $x \in \mathbb{R}^{N\times 1}$ is the image, $D \in \mathbb{R}^{MK\times K}$ is the direct arrival signal modeling matrix, $g \in \mathbb{R}^{K\times 1}$ is a vector containing scaling coefficients for the direct arrival signals, $M$ is the number of measurement samples, and $N$ is the number of pixels.
The columns of $D$, $d_k$, are the discretized version of $\tilde{d}_k$.
The vector $g$ is used to scale each column of $D$ independently.

\subsection{Joint-MAP Stitching}
\begin{figure}
	\centering
	\includegraphics[width=0.4\textwidth]{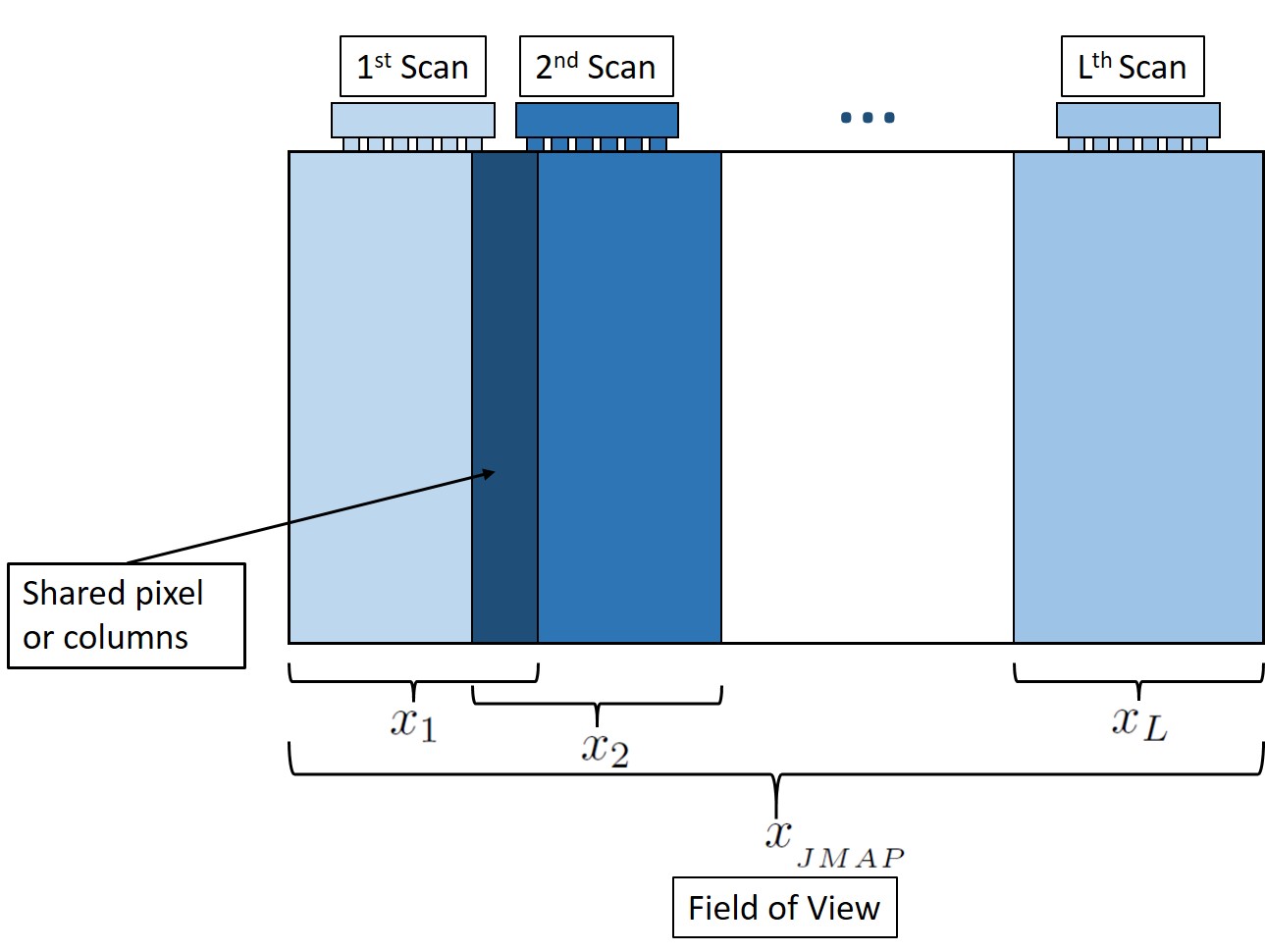}
	\caption{An illustration of multiple measurements needed to scan a large field-of-view. Images from each scan share some pixels with its neighbor images. Proper stitching technique is needed to account for this shared areas in the field-of-view.}
	\label{fig:joint_map}
\end{figure}

In order to scan large regions, the sensor assembly is typically moved from one region to another on the surface in raster order to build up a 3D profile of the strcuture. Typically each data set is individually processed and placed together to present the overall 3D reconstruction, Fig. \ref{fig:joint_map}. 
However, this method results in sharp discontinuities at the boundaries and inefficient use of the data collected, Fig. \ref{fig:illu_SAFT}.
We design a joint-MAP technique to solve these issues by modifying the forward model to perform the stitching internally as part of the estimation.
This technique is able to remove all discontinuities between the sections, make use of any additional information from adjacent scans, and process each pixel in the large field-of-view once.
We assume that adjacent scans share some columns of pixels and has some useful correlation that needs to be exploited to produce better images.
Therefore, the forward model will account for those shared columns differently than the rest of the pixels or columns.
For $L$ measurements, we let the system matrix for each measurement be $A$ and the image for each measurement be $x_l$. 
We let the order of the pixels in $x_l$ be from top to bottom for each column starting from the far left column to the far right column.
Hence, the term associated with the modified forward model in the MAP estimate will be
\begin{eqnarray}
\frac{1}{2\sigma^2} \left\|  y_{ _{ _{JMAP}}}-A_{ _{ _{JMAP}}}x_{ _{ _{JMAP}}} - D_{ _{ _{JMAP}}}g_{ _{ _{JMAP}}}\right\|^2,
\end{eqnarray}
where
\begin{eqnarray*}
A_{ _{ _{JMAP}}} &=& 
\begin{bmatrix}
[& &A& &]&0& & &0& & &\dots\\
 & &0&[& &A& &]&0& & &\dots\\
 & &0& & &0&[& &A& &]&\dots\\
 & &\vdots& & &\vdots& & &\vdots& & &\ddots
\end{bmatrix},\\
\\
D_{ _{ _{JMAP}}} &=&  \begin{bmatrix}
	D & 0 & \hdots & 0\\ 
	0 & D & \hdots & 0\\ 
	\vdots & \vdots & \ddots & \vdots\\ 
	0 & 0 & \hdots & D
\end{bmatrix},\\
y_{ _{ _{JMAP}}} &=& \begin{bmatrix}
	y_1\\
	\vdots\\
	y_l\\
	\vdots\\
	y_L
\end{bmatrix}, \quad
g_{ _{ _{JMAP}}} = \begin{bmatrix}
g_1\\
\vdots\\
g_l\\
\vdots\\
g_L
\end{bmatrix},
\end{eqnarray*}
and $x_{ _{ _{JMAP}}}$ is the image of the large field-of-view.
$A_{ _{ _{JMAP}}}$ is designed so that if a pixel is shared in more than one image, then its corresponding column in the system matrix for one image will be aligned with its corresponding columns in the system matrix for other images. 
For the example shown in Fig. \ref{fig:joint_map}, we can accomplish this alignment by shifting each system matrix $A$ left or right until the required alignment is achieved.
\section{Prior Model Of the image}
\label{prior_model}
We design the forward model of the image to be a combination of a Gibbs and an exponential distribution, i.e.
\begin{eqnarray*}
-\log p(x) = \sum_{\{s,r\} \in C} b_{s,r} \ \rho(x_s-x_r,\sigma_g) + \sum_{s \in S} \frac{x_s}{\sigma_{e}}  + \text{constant}, 
\end{eqnarray*}
where $C$ is the set of all pair-wise cliques, $S$ is the set of all pixels in the field of view, $b_{s,r}$ is a scaling coefficient, $\rho$ is the potential function, $\sigma_g$ is the regularization constant for the Gibbs distribution, $\sigma_{e}$ is the regularization constants for the exponential distribution, and $x_s \geq 0 \ \forall s \in S$ .
We chose the q-generalized Gaussian Markov random field (QGGMRF) as the potential function for the Gibbs distribution \cite{BoumanBook2013}.
The equation for the QGGMRF is 
\begin{eqnarray}
	\rho(\Delta,\sigma_g) = \frac{|\Delta|^p}{p\sigma_g^p}\left( \frac{|\frac{\Delta}{T\sigma_g}|^{q-p}}{1+|\frac{\Delta}{T\sigma_g}|^{q-p}}\right), \label{QGGMRF}
\end{eqnarray}
where $1\le p < q = 2$ insures convexity and continuity of first and second derivatives, and $T$ controls the edge threshold.

The neighbors of a pixel $s$ are arranged as 
\begin{eqnarray}
\begin{bmatrix}
r_1&r_2&r_3\\
r_4&r_5&r_6\\
r_7&r_8&r_9
\end{bmatrix},
\begin{bmatrix}
r_{10}&r_{11}&r_{12}\\
r_{13}&s&r_{14}\\
r_{15}&r_{16}&r_{17}
\end{bmatrix},
\begin{bmatrix}
r_{18}&r_{19}&r_{20}\\
r_{21}&r_{22}&r_{23}\\
r_{24}&r_{25}&r_{26}
\end{bmatrix}. \label{arrang}
\end{eqnarray}
where the neighbors with index 10 to 17 are from the same layer, and the rest of the neighbors are from the next and previous layers.
With this arrangement, the scaling coefficients $b_{s,r}$ are chosen to be
\begin{eqnarray*}
	\begin{bmatrix}
		b_{s,r_{1}}&b_{s,r_{2}}&b_{s,r_{3}}\\
		b_{s,r_{4}}&b_{s,r_{5}}&b_{s,r_{6}}\\
		b_{s,r_{7}}&b_{s,r_{8}}&b_{s,r_{9}}
	\end{bmatrix} &=& \begin{bmatrix}
		0&0&0\\
		0&2&0\\
		0&0&0
	\end{bmatrix}\cdot \frac{\gamma}{4\gamma+12},\\
	\begin{bmatrix}
		b_{s,r_{10}}&b_{s,r_{11}}&b_{s,r_{12}}\\
		b_{s,r_{13}}&0&b_{s,r_{14}}\\
		b_{s,r_{15}}&b_{s,r_{16}}&b_{s,r_{17}}
	\end{bmatrix} &=& \begin{bmatrix}
		1&2&1\\
		2&0&2\\
		1&2&1
	\end{bmatrix}\cdot \frac{1}{4\gamma+12},\\
	\begin{bmatrix}
		b_{s,r_{18}}&b_{s,r_{19}}&b_{s,r_{20}}\\
		b_{s,r_{21}}&b_{s,r_{22}}&b_{s,r_{23}}\\
		b_{s,r_{24}}&b_{s,r_{25}}&b_{s,r_{26}}
	\end{bmatrix} &=& \begin{bmatrix}
		0&0&0\\
		0&2&0\\
		0&0&0
	\end{bmatrix}\cdot \frac{\gamma}{4\gamma+12},\label{bsr}
\end{eqnarray*}
with a free boundary condition.
The parameter $\gamma$ is set to zero when 2D MBIR is needed, or greater than zero when a 3D regularization (2.5D MBIR) is needed.
2.5D MBIR can be used to gain more information from neighbors of different layers to reduce noise and increase resolution.
\subsection{Non-linear Spatially-Variant Regularization}

\begin{figure}
	\centering
	\includegraphics[width=0.3\textwidth]{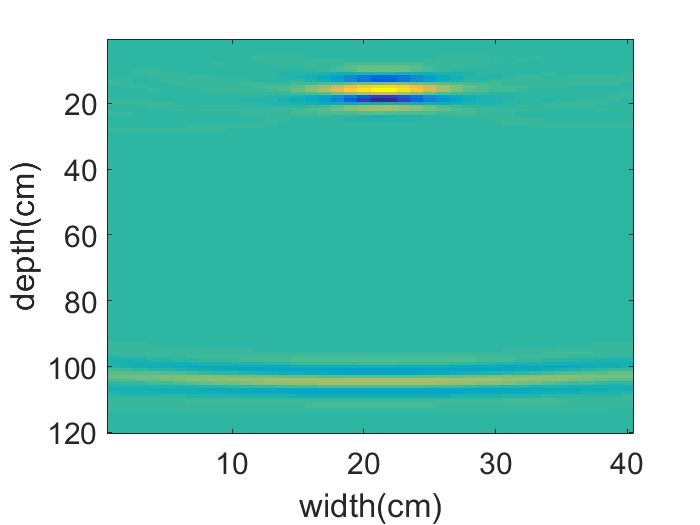}
	\caption{Back-projection of two point scatters, one that is closer to the transducers (17cm deep) and one that is far from the transducers (105 cm deep). As the reflection gets deeper, The lateral resolution decreases.}
	\label{bachprojection} 
\end{figure}

The standard form of the regularization introduced above uses constant $\sigma_g$ and $\sigma_e$ for all voxels. 
However, this can result in reconstruction artifacts because for closer reflections, there are few pixels that could have contributed to the signal. 
However, for deeper reflections, there are many more pixels that could have caused the reflection, i.e. the deeper the reflection the less lateral resolution it has. 
Fig. \ref{bachprojection} shows the back-projection of two point scatters of different depth. 
The closer reflection has less overlapping and higher lateral resolution.
The deeper reflection has larger overlapping and lower lateral resolution.
This is an issue because MBIR spreads the energy over the intersection area, which attenuates the intensity dramatically for deeper reflections.
This smoothing and attenuation appear to increase more rapidly for deeper reflection.
Therefore, a linear spatially-variant regularization as in \cite{ozkan2018inverse} is not sufficient, and a more generalized model is needed. 
Hence, we adapt a non-linear spatially-variant regularization technique designed for the UNDE system.
We can solve the attenuation problem by assigning less regularization as the pixel gets deeper. 
The disadvantage of this method is that it will amplify both the reflection and the noise for deeper pixels. 

We replace $\sigma_g$ and $\sigma_e$ with $\sigma_{g_{s,r}}$ and $\sigma_{e_{s}}$, respectively, where these new parameters are monotone increasing with respect to depth. 
We assign a new scaling parameter $c_s$ that varies between two values $c_{s_{\text{min}}}$ and $c_{s_{\text{max}}}$ as follows:
\begin{eqnarray}
c_s = c_{s_{\text{min}}} + (c_{s_{\text{max}}}-c_{s_{\text{min}}})*\left( \frac{\text{depth}}{\text{maximum depth}}\right) ^a \label{spatial}
\end{eqnarray}
where $a>0$.
Then, $\sigma_{g_{s,r}}$ and $\sigma_{e_{s}}$ are calculated as follows:
\begin{eqnarray*}
	\sigma_{g_{s,r}} &=& \sigma_g \sqrt{c_sc_r},\\
	\sigma_{e_s} &=& \sigma_e c_s \ .
\end{eqnarray*}
\section{Optimization of MAP Cost Function}
\label{Opt}
\begin{figure}
	\begin{center}
		\begin{tabular}{c}
			\includegraphics[width=0.25\textwidth]{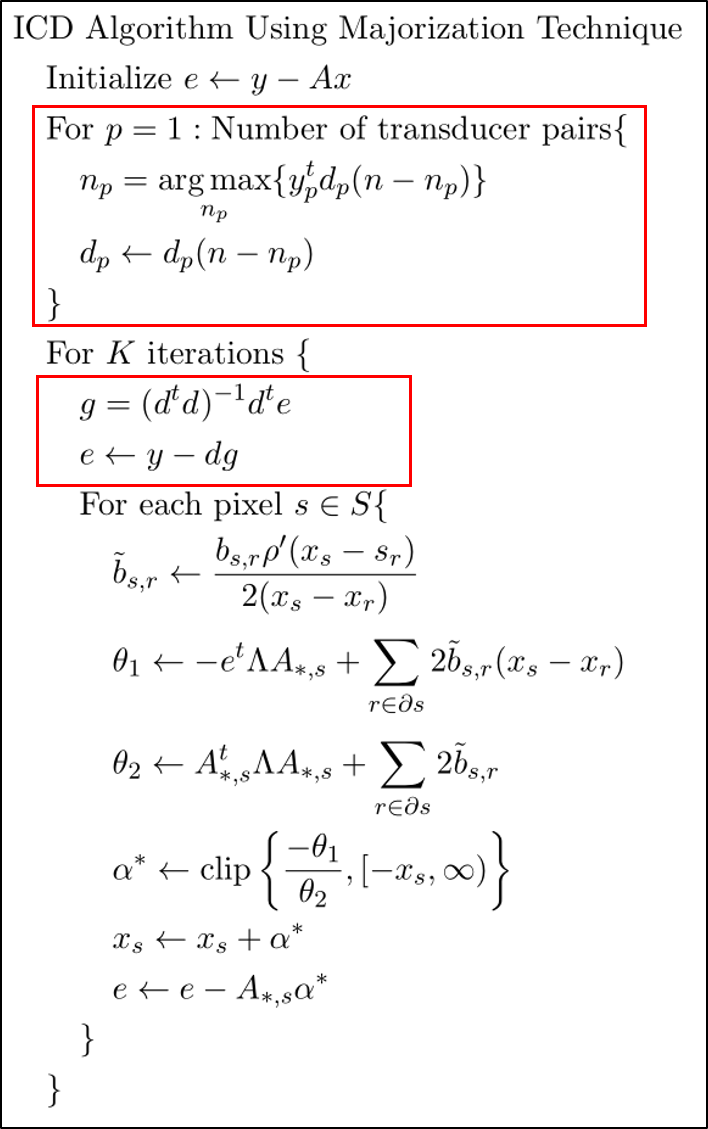}
		\end{tabular}
	\end{center}
	\vspace{-0.15in}
	\caption{\label{DIRECTSIGNAL}
		\scriptsize
		{ICD algorithm using the majorzation technique with shift error estimation (top red box) and direct arrival modeling (bottom red box)\cite{BoumanBook2013,yu2011fast}.}}
\end{figure}

After designing the forward model and the prior model, the MAP cost function is
\begin{eqnarray}
	\begin{split}
		(x,g)_{MAP} &= \underset{x \geq 0,g} {\text{arg\,min}} \biggl\lbrace \frac{1}{2\sigma^2} \left\|  y-Ax-Dg\right\|^2\\
		&+ \sum_{\{s,r\} \in C} b_{s,r} \ \rho(x_s-x_r,\sigma_{g_{s,r}}) + \sum_{s \in S} \frac{x_s}{\sigma_{e_{s}}}\biggl\rbrace.
	\end{split} \label{map}
\end{eqnarray}
The shifting of the direct arrival signal matrix $D$ mentioned in section \ref{shift_error} is performed once before evaluating $x$ and $g$.
The solution for $g$ is straightforward:
\begin{eqnarray*}
	&&\begin{split}
		0 &=\bigtriangledown_{g} \biggl\lbrace \frac{1}{2\sigma^2} \left\|  y-Ax-Dg\right\|^2\\
		& + \sum_{\{s,r\} \in C} b_{s,r} \ \rho(x_s-x_r) \biggl\rbrace\\
		\implies 0 &= 2D^tDg+2D^tAx-2D^ty\\
		\implies g &= (D^tD)^{-1}D^t(y-Ax).
	\end{split}
\end{eqnarray*}
Given $x$, the evaluation of $g$ is computationally inexpensive because $D^tD$ is a diagonal matrix, i.e.
\begin{eqnarray*}
	(D^tD) = \begin{bmatrix}
		d_1^td_1^{} & 0 & \hdots & 0\\ 
		0 & d_2^td_2^{} & \hdots & 0\\ 
		\vdots & \vdots & \ddots & \vdots\\ 
		0 & 0 & \hdots & d_K^td_K^{}
	\end{bmatrix},
\end{eqnarray*}
where $K$ is the total number of transducer pairs and $d_k$ is the discretized version of $\tilde{d}_k(t)$ for transducer pair $k$.
However, $g$ requires the knowledge of $x$ which is the image we would like to reconstruct.
This issue can be resolved by updating the value of $g$ from the updated image in each iteration. 
Furthermore, for each iteration, we update $g$ and $x$ in the following steps:
\begin{eqnarray*}
g &\leftarrow& (D^tD)^{-1}D^t(y-Ax)\\
y &\leftarrow& y - Dg\\
x &\leftarrow& \underset{x \geq 0} {\text{arg\,min}} \biggl\lbrace -\log p(y|x) -\log p(x) \biggl\rbrace
\end{eqnarray*}
We adopt the iterative coordinate descent (ICD) technique to optimize the cost function with respect to $x$ \cite{yu2011fast}. 
Since the prior model term is non-quadratic, optimizing the cost function will be computationally expensive.
Therefore, we use the surrogate function (majorization) approach with ICD to resolve this issue \cite{BoumanBook2013}. 
Fig. \ref{DIRECTSIGNAL} shows the complete algorithm for ICD using the majorzation approach.

\section{Results}

\subsection{K-wave Simulated Results}
\begin{figure*}
	\begin{center}
		\begin{tabular}{@{}c@{}c@{}c@{}c@{}c@{}}
			&Defect Diagram&SAFT&$l_1$-norm&MBIR
			\tabularnewline
			\rot{Test 1}&
			\includegraphics[align = c,width=0.2\textwidth]{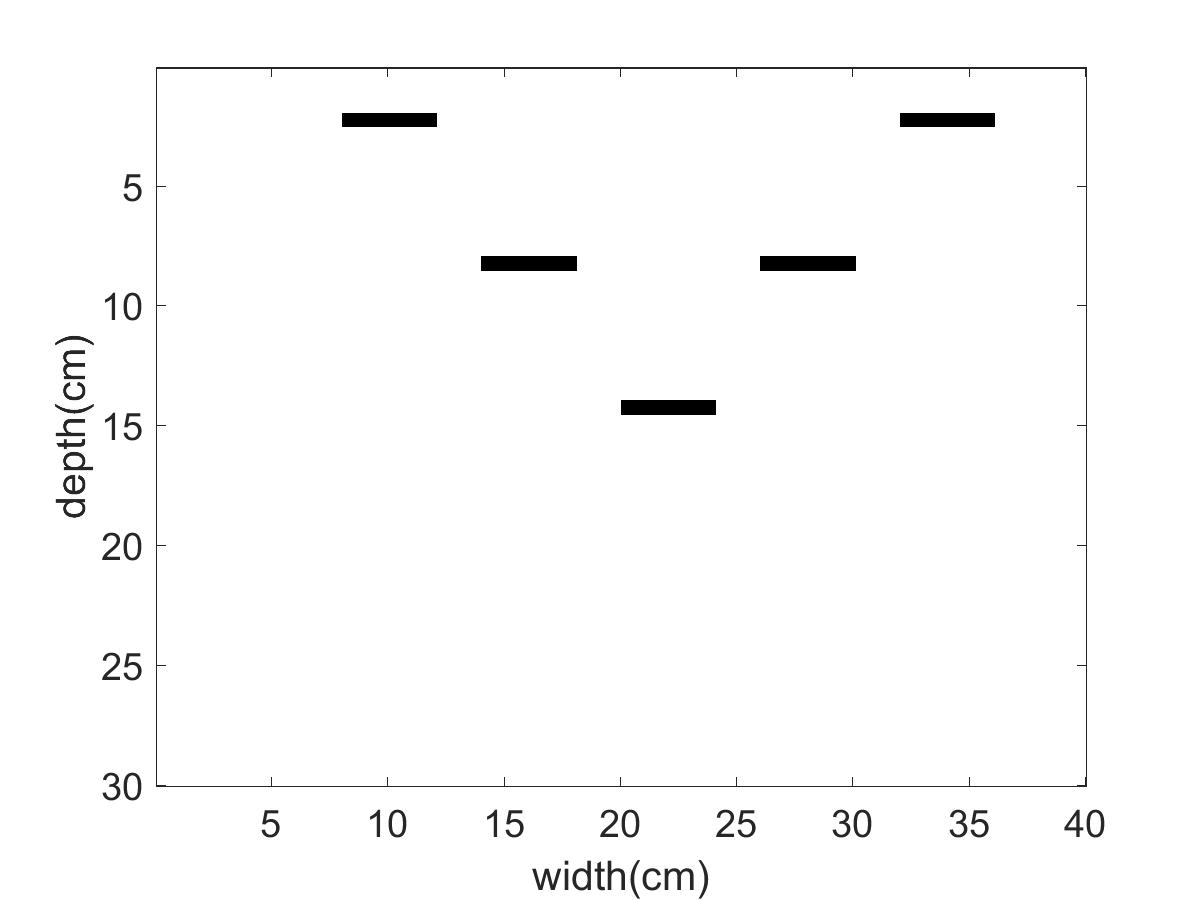} &
			\includegraphics[align = c,width=0.2\textwidth]{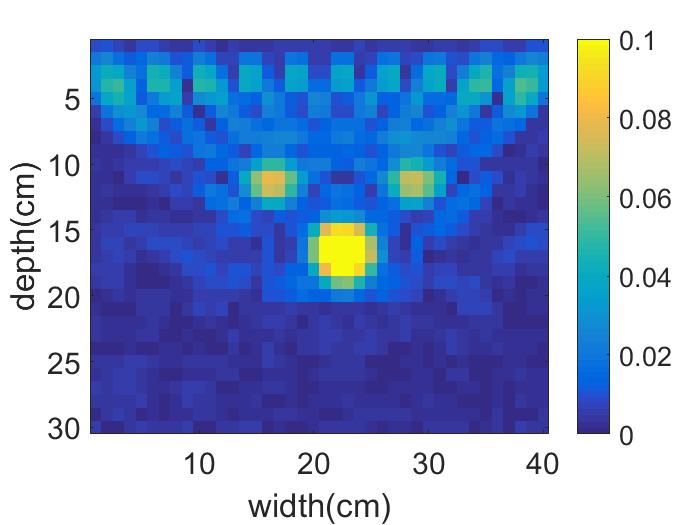}&
			\includegraphics[align = c,width=0.2\textwidth]{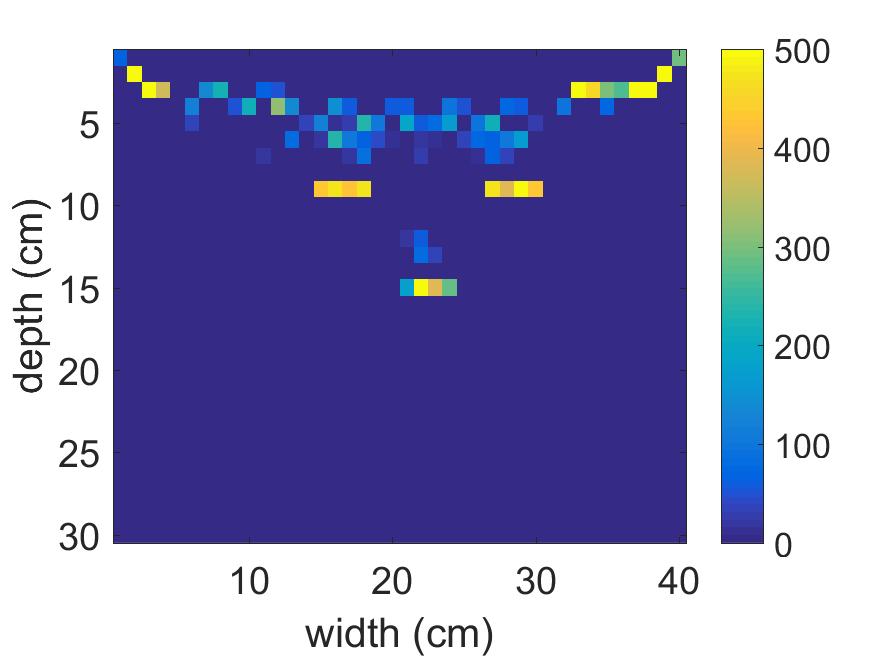}&
			\includegraphics[align = c,width=0.2\textwidth]{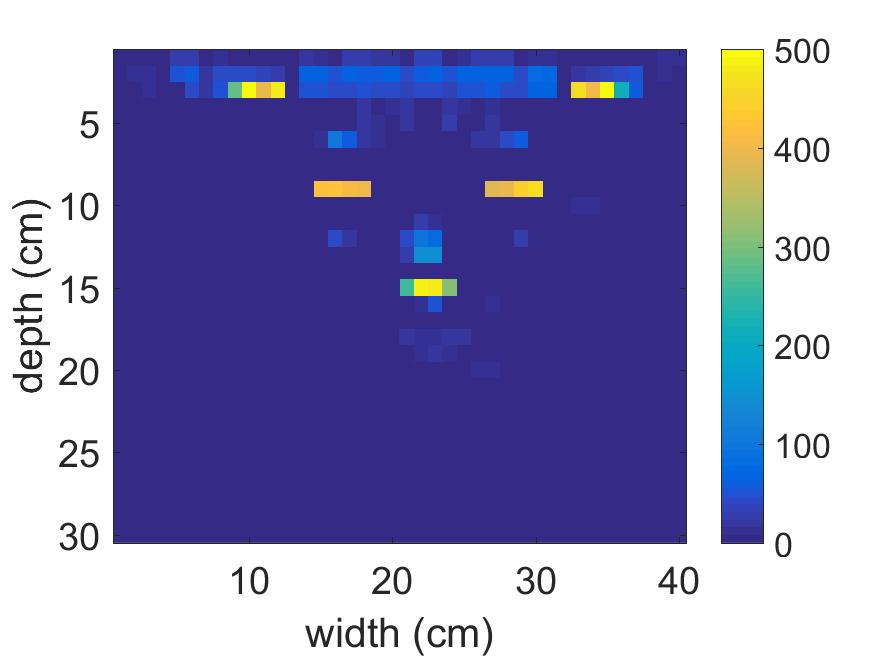}
			\tabularnewline
			\rot{Test 2} &
			\includegraphics[align = c,width=0.2\textwidth]{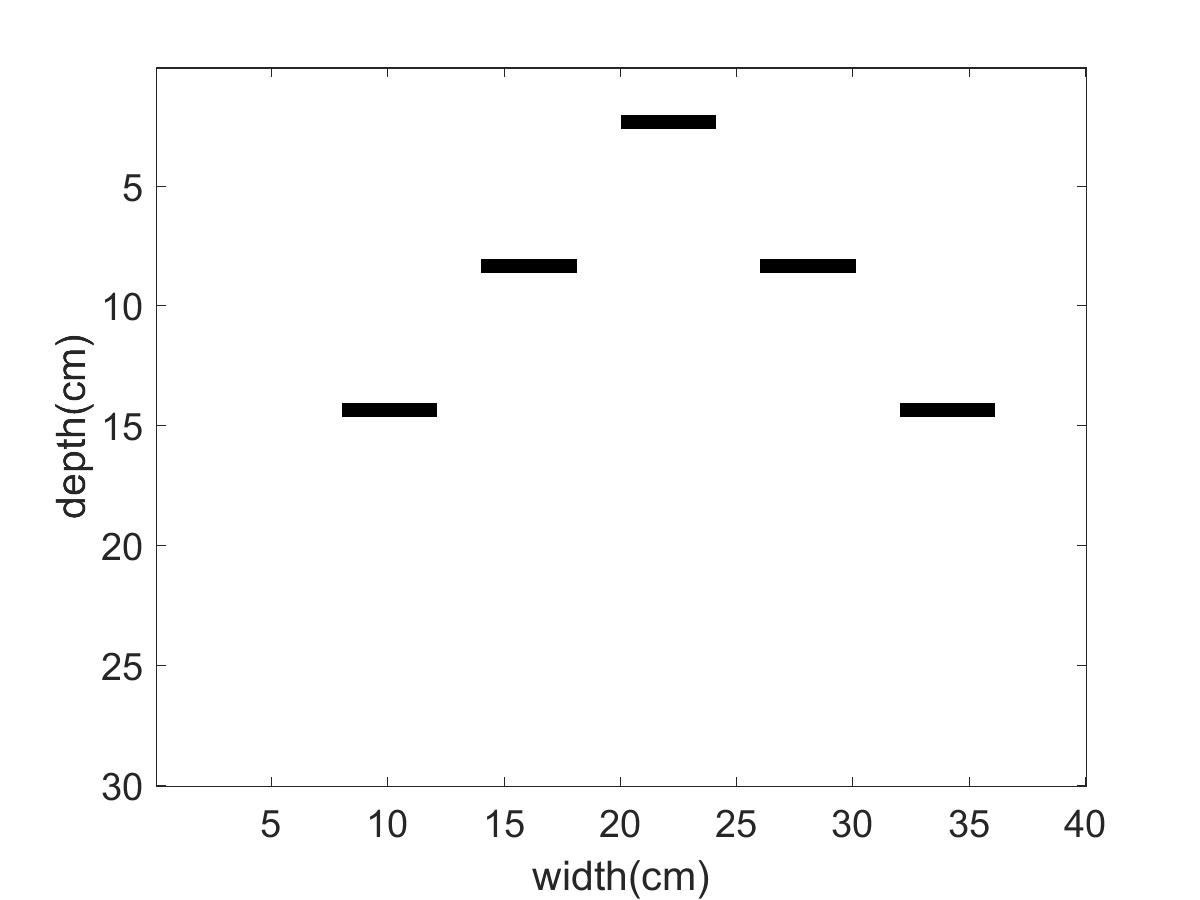} &
			\includegraphics[align = c,width=0.2\textwidth]{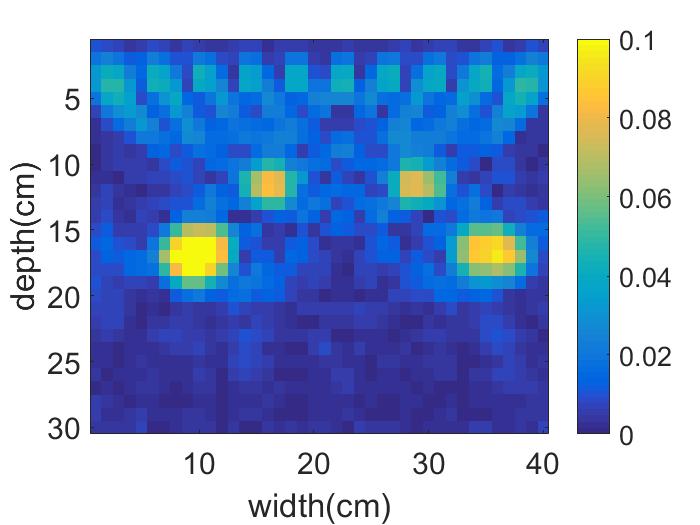}&
			\includegraphics[align = c,width=0.2\textwidth]{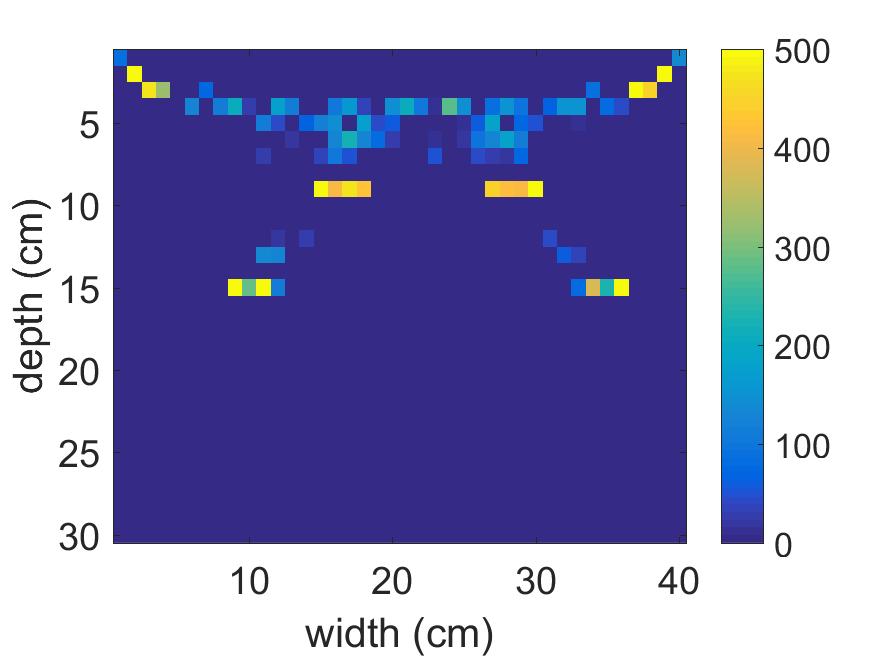}&
			\includegraphics[align = c,width=0.2\textwidth]{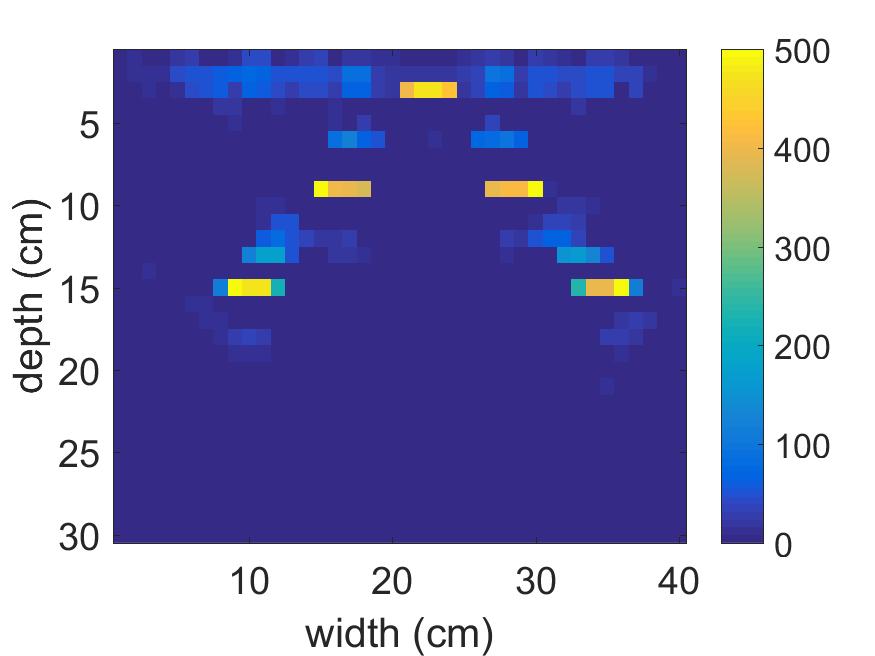}
			\tabularnewline
			\rot{Test 3} &
			\includegraphics[align = c,width=0.2\textwidth]{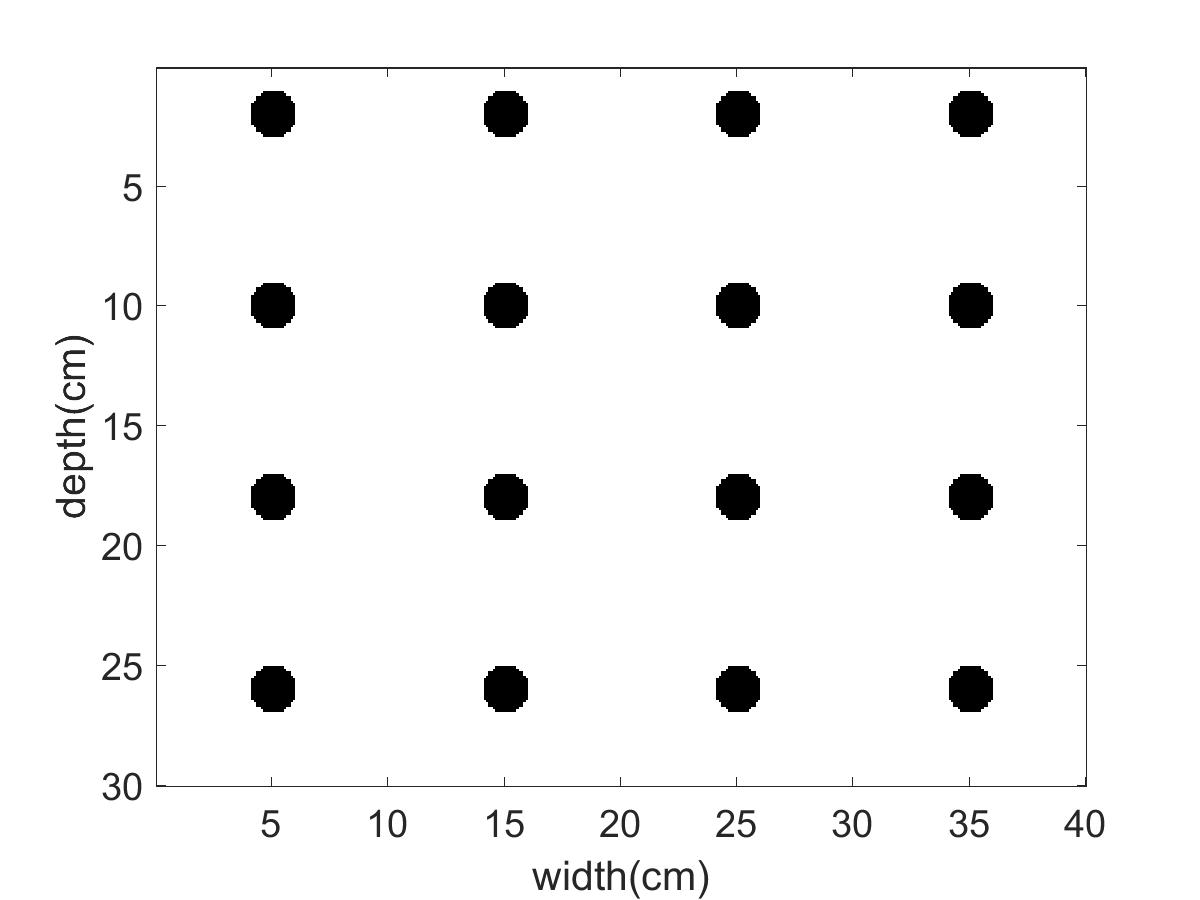} &
			\includegraphics[align = c,width=0.2\textwidth]{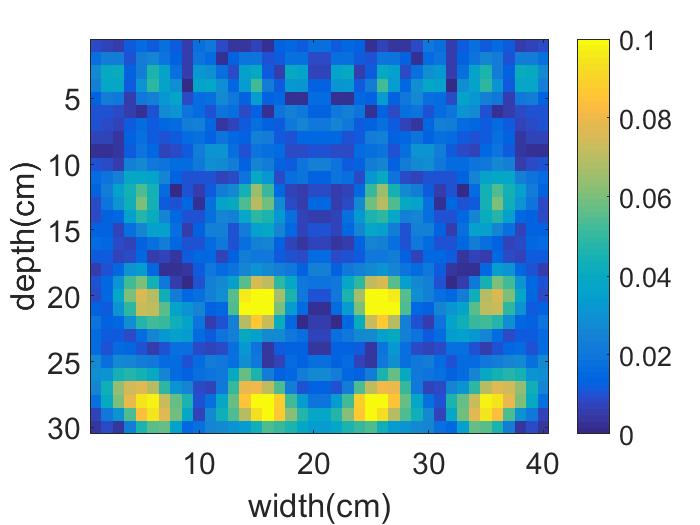}&
			\includegraphics[align = c,width=0.2\textwidth]{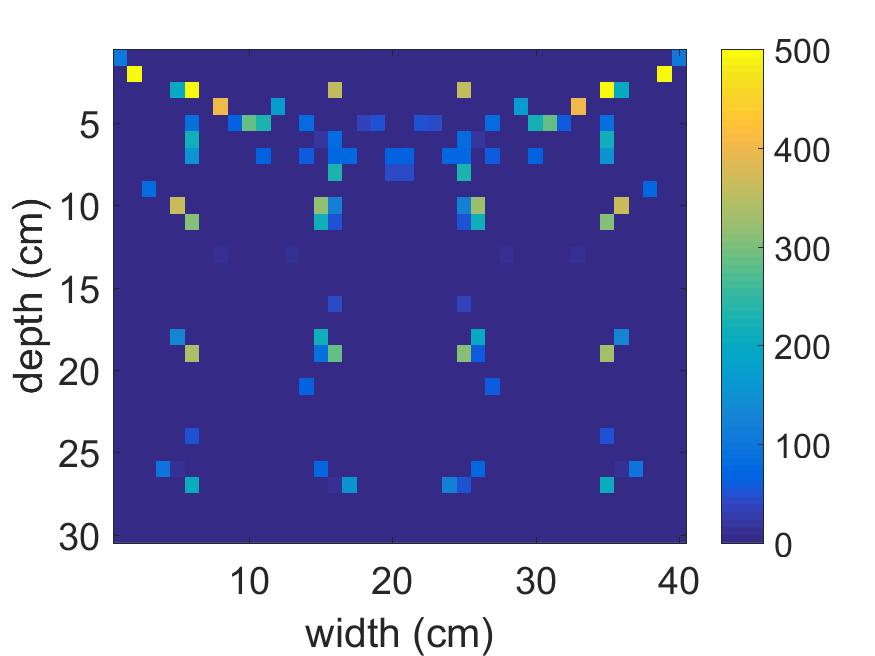}&
			\includegraphics[align = c,width=0.2\textwidth]{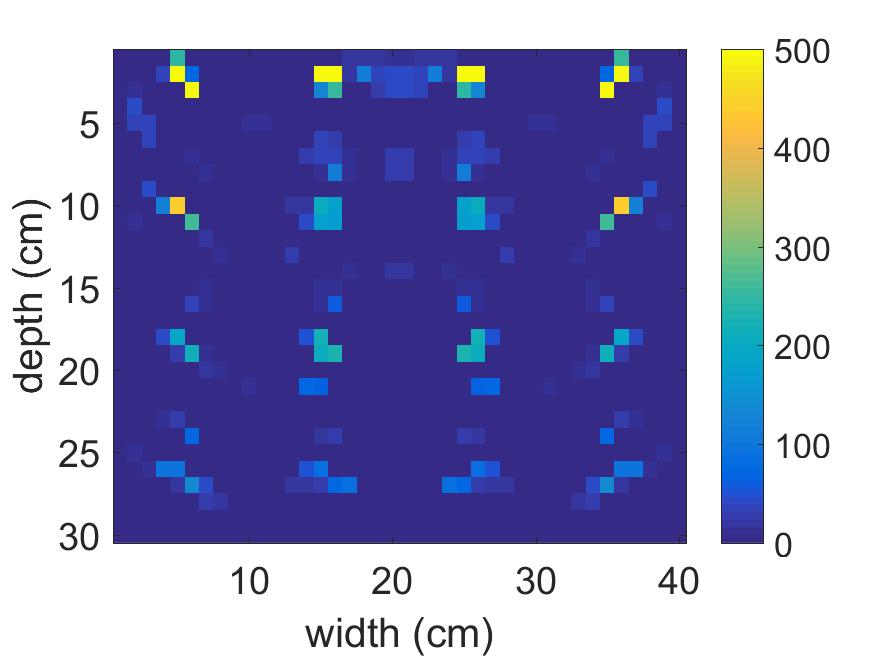}
			\tabularnewline
			\rot{Test 4} &
			\includegraphics[align = c,width=0.2\textwidth]{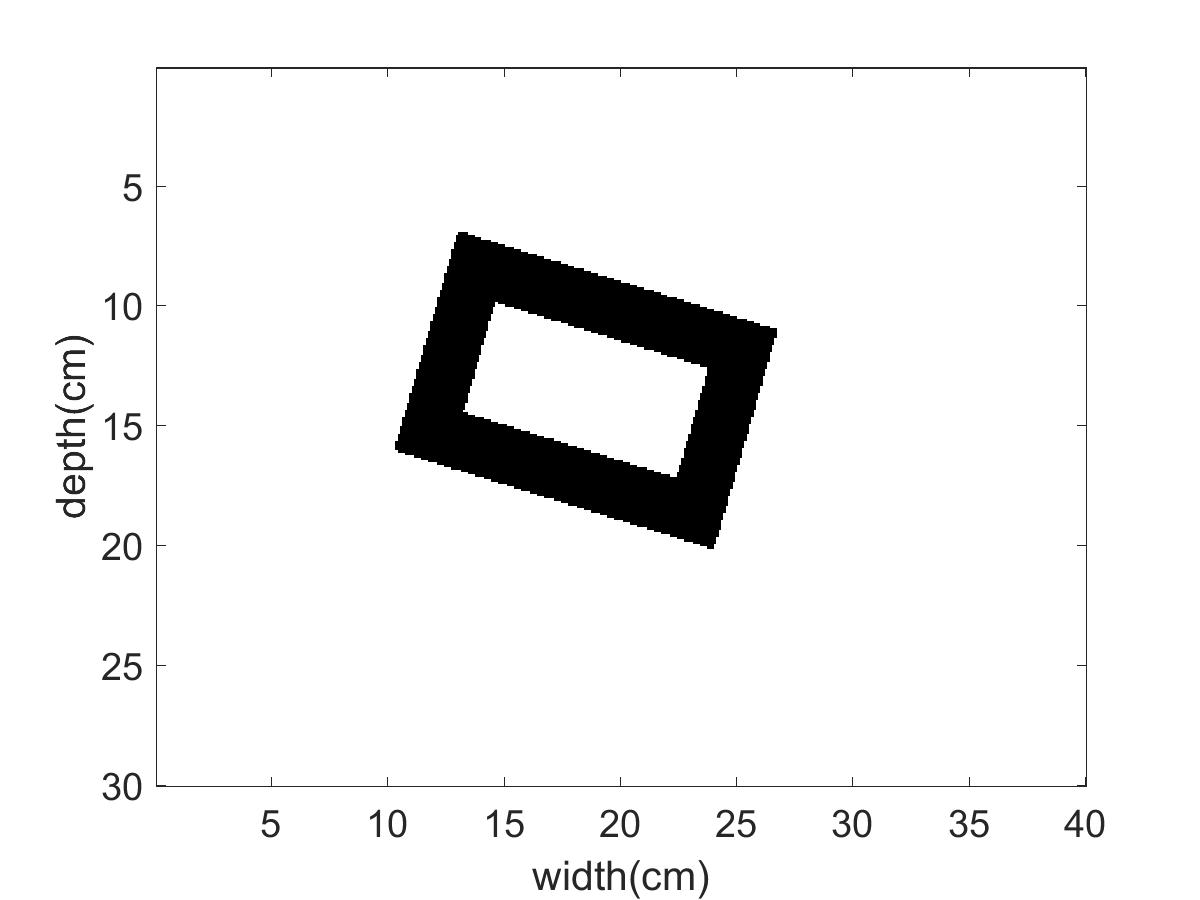} &
			\includegraphics[align = c,width=0.2\textwidth]{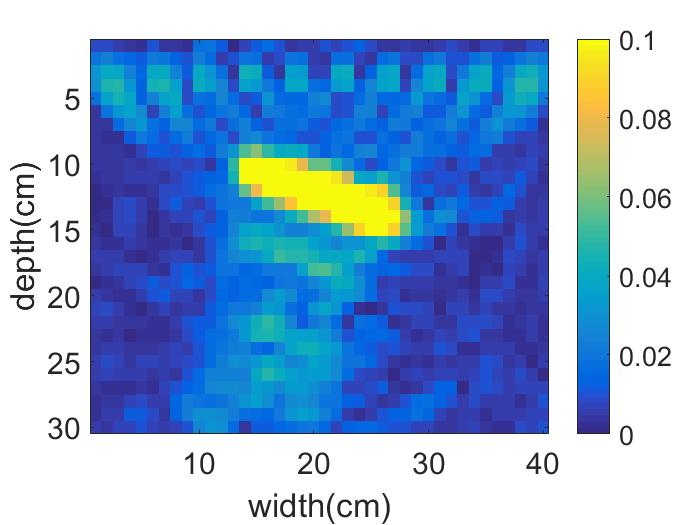}&
			\includegraphics[align = c,width=0.2\textwidth]{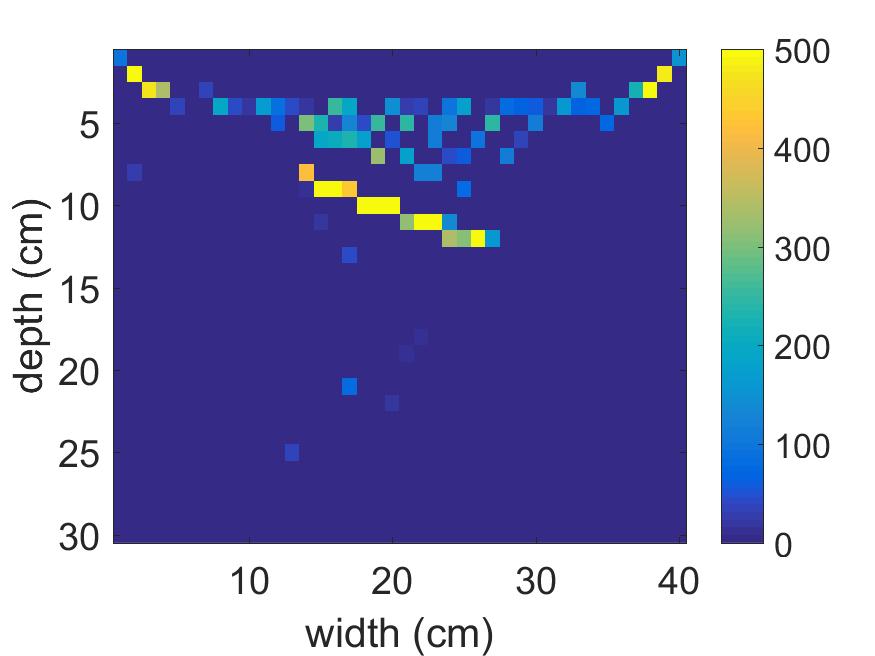}&
			\includegraphics[align = c,width=0.2\textwidth]{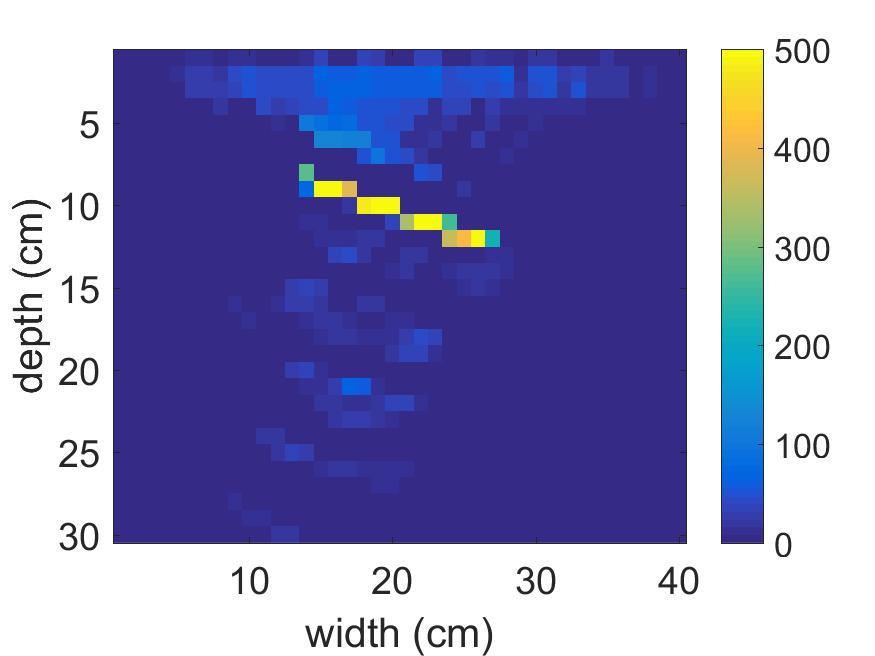}
			\tabularnewline
			
		\end{tabular}
	\end{center}
	\vspace{-0.15in}
	\caption{\label{k-wave}
		\scriptsize
		{Comparison between MBIR and SAFT reconstruction from the k-wave simulated data. The far left column is the position of the defects. The next column is SAFT reconstruction. The next column is $l_1$-norm reconstruction. The far right column is MBIR reconstruction. MBIR tends to produce results with less noise and artifacts compared to SAFT and $l_1$-norm.}}
\end{figure*}

The k-wave simulator have been used to simulate acoustic propagation through concrete medium \cite{treeby2010k}. 
The concrete structure was embedded with steel of different shapes.
The width and depth of the structure is 40cm and 30cm , respectively.
10 transducers were used to transmit and receive.
For each simulation, the simulator produces 90 outputs from all pairs of transducers where only distinctive pairs are used, i.e. 45 distinctive pairs.
The transducers are placed at the top center of the field-of-view and separated by 4cm from each other.
To simulate the acoustic propagation using k-wave, we provided three images of speed, density, and attenuation as inputs to k-wave.
Each pixel in the three input images corresponds to the characteristics of either steel or cement.
The output of k-wave is then used as input to the reconstruction methods.
Fig. \ref{k-wave} shows reconstruction results for four different tests.
The voxel spacing for 2D reconstructions is 1 cm for all reconstruction techniques.
The left column shows the designed defect diagram that was used for simulation where the white pixels corresponds to cement and the black pixels corresponds to steel.
The next column shows the instantaneous envelope of SAFT reconstruction.
The next column is a regularized iterative technique with the same forward model as in Eq. \ref{forwardmodel2} with an exponential distribution prior.
The prior model is exactly equal to an $l_1$ regularization term with a positivity constraint.
This technique will be referred to as $l_1$-norm for the rest of the paper.  
The MAP estimate for the $l_1$-norm technique is 
\begin{eqnarray}
	x_{MAP} = \underset{x \geq 0} {\text{arg\,min}} \biggl\lbrace \frac{1}{2\sigma^2} \left\|  y-Ax\right\|^2 +\sum_{s \in S} \frac{x_s}{\sigma_{e_{s}}} \biggl\rbrace. \label{l1norm}
\end{eqnarray}
The right column shows the MBIR reconstruction.
Note that SAFT does not share the same unit with MBIR or L1-norm. That is why it shows different scaling.

\begin{figure*}
	\begin{center}
		\begin{tabular}{@{}c@{}c@{}c@{}c@{}}
			&SAFT&$l_1$-norm&MBIR
			\tabularnewline
			\rot{SNR = 3} \ &
			\includegraphics[align = c,width=0.2\textwidth]{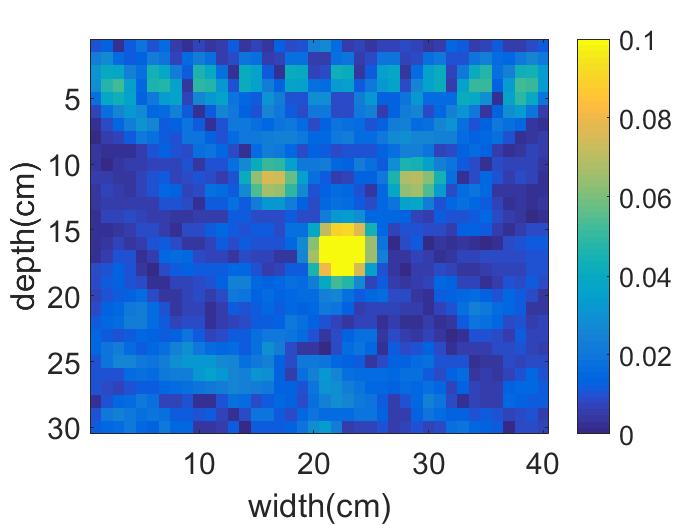}&
			\includegraphics[align = c,width=0.2\textwidth]{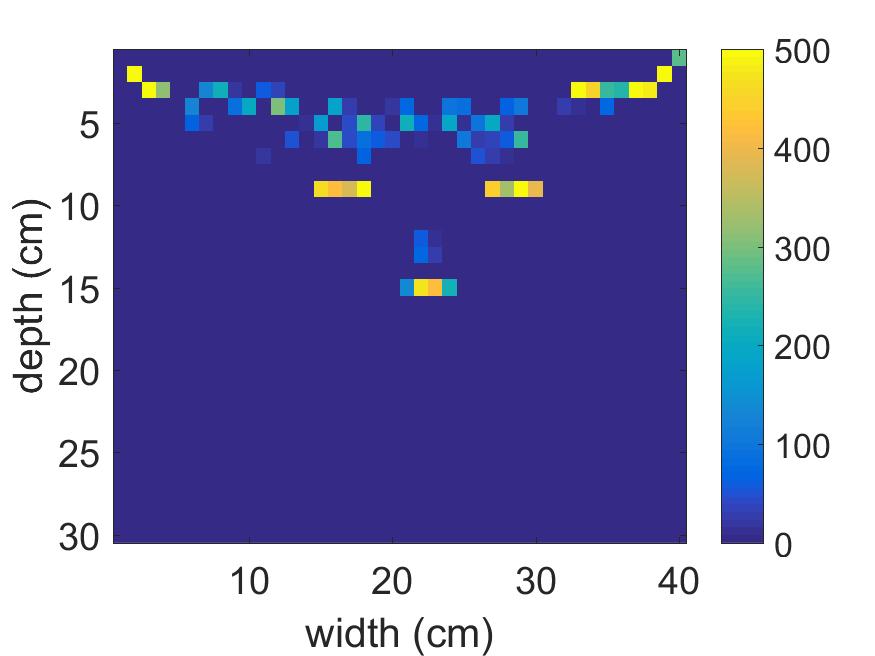}&
			\includegraphics[align = c,width=0.2\textwidth]{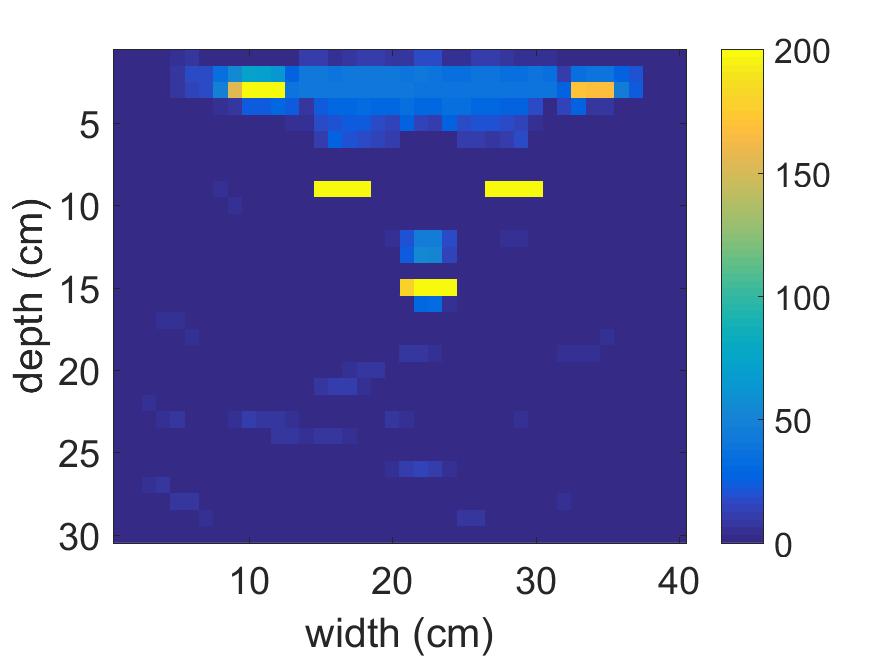}
			\tabularnewline
			\rot{SNR = 1} \ &
			\includegraphics[align = c,width=0.2\textwidth]{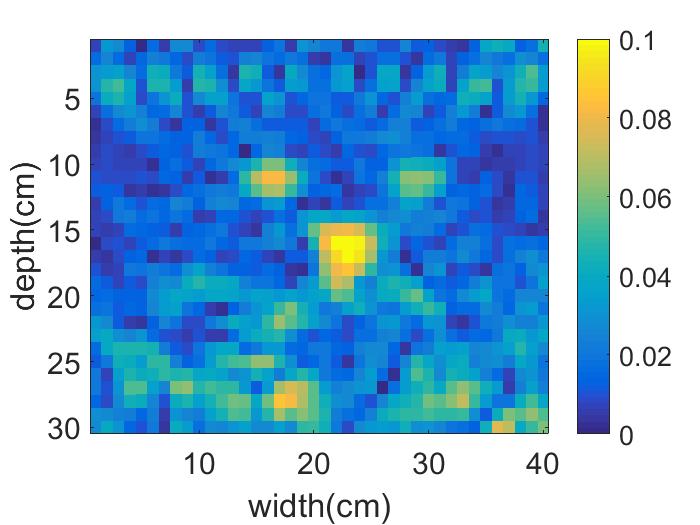}&
			\includegraphics[align = c,width=0.2\textwidth]{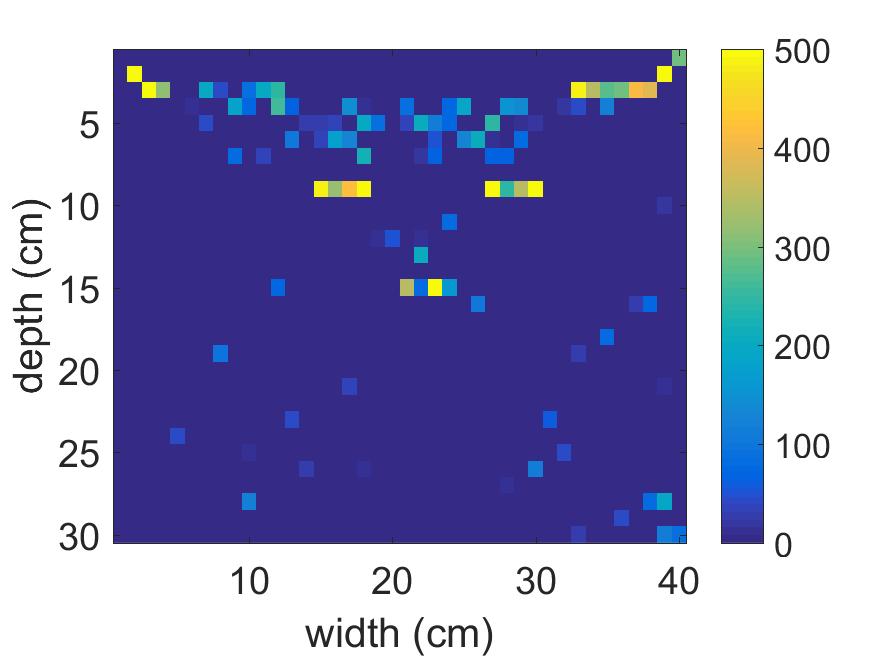}&
			\includegraphics[align = c,width=0.2\textwidth]{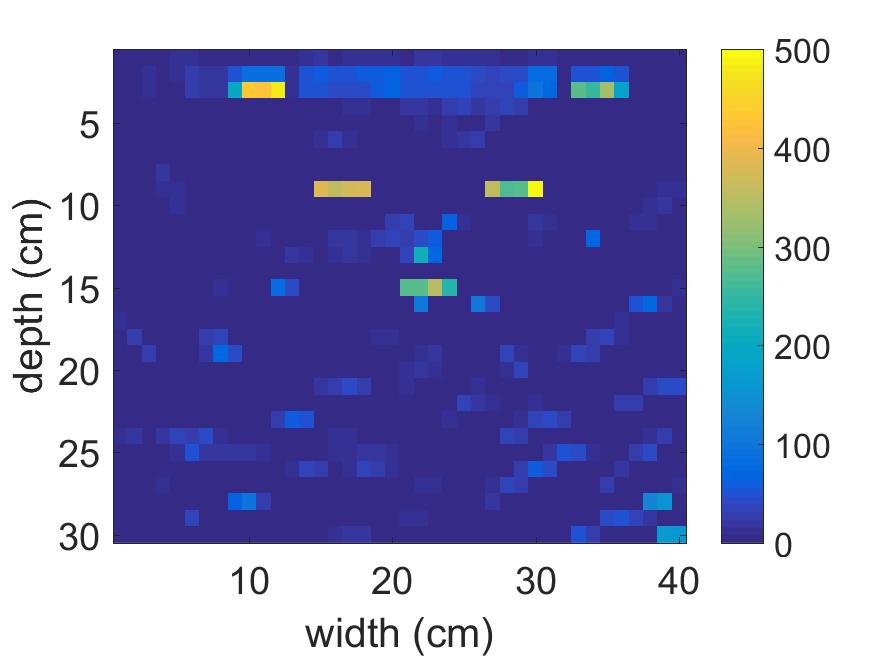}
			\tabularnewline
			\rot{SNR = 0.33} \ &
			\includegraphics[align = c,width=0.2\textwidth]{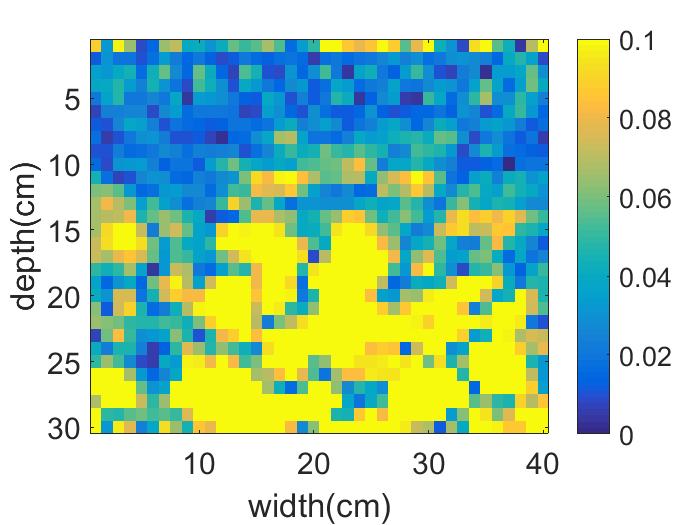}&
			\includegraphics[align = c,width=0.2\textwidth]{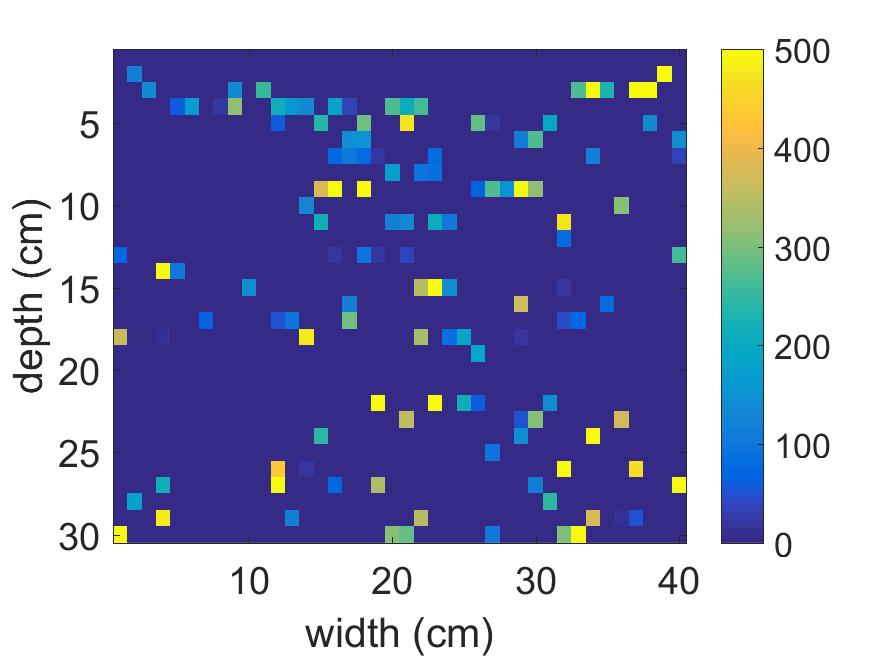}&
			\includegraphics[align = c,width=0.2\textwidth]{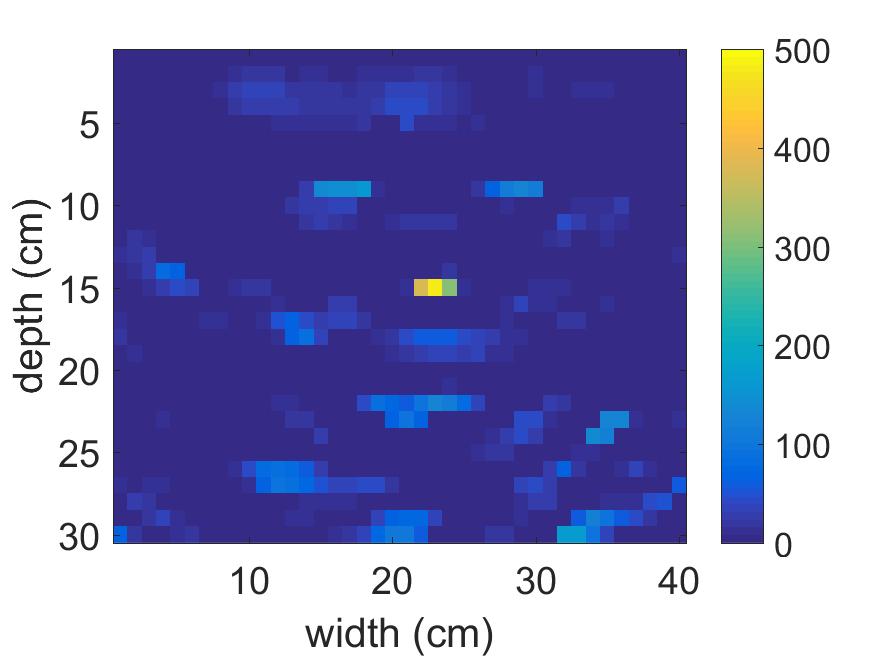}
			\tabularnewline
			
		\end{tabular}
	\end{center}
	\vspace{-0.15in}
	\caption{\label{k-wave_noisy}
		\scriptsize
		{Comparison between SAFT, $l_1$-norm, MBIR reconstructions from the k-wave simulated data with different SNR. The defect diagram is the same as the defect diagram in Test 1 in Fig. \ref{k-wave}. The left column is SAFT reconstruction. The next column is $l_1$-norm reconstruction. The right column is MBIR reconstruction. Each row correspond to different SNR value where the SNR values from top to bottom are 3, 1, and 0.33, repectively. MBIR tends to produce results with less noise and artifacts compared to SAFT and $l_1$-norm.}}
\end{figure*}

\begin{figure}
	\centering
	\includegraphics[width=0.4\textwidth]{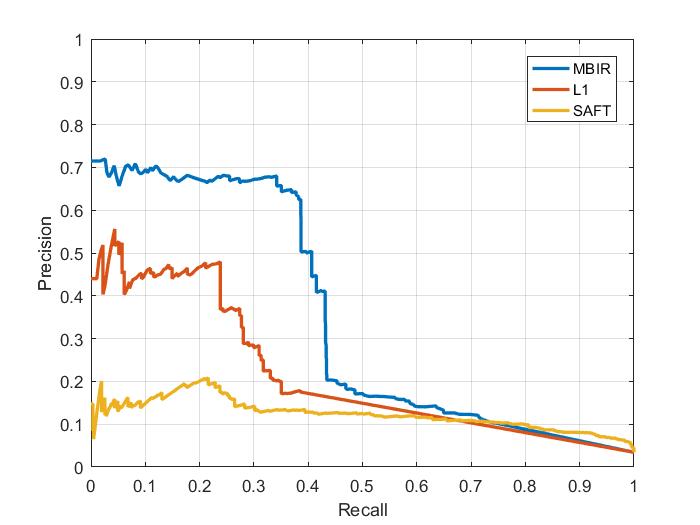}
	\caption{PR curves for each technique over all 4 tests in Fig.\ref{k-wave}. MBIR outperforms the other techniques by having the highest PR area.}
	\label{fig:pr_k_wave}
\end{figure}

A pixel-wise detection test was performed for all 4 tests to calculate the number of true positive (TP), false positive (FP), and false negative (FN) for each technique.
These values are, then, used to plot the precision vs. recall (PR) curves where 
$$
recall = \frac{TP}{TP+FN}
$$
and
$$
precision = \frac{TP}{TP+FP}.
$$
This detection test compares the performance of each technique by the area under the PR curve.
The larger the area the better the technique.
Next, for each technique, all the images are normalized by dividing them with their maximum value.
Thresholds from 1 to 0 with step 0.001 are applied to all images.
For each threshold, a TP is declared if the defect diagram pixel is 1 and the reconstructed pixel is 1.
A FP is declared if the defect diagram pixel is 0 and the reconstructed pixel is 1.
A FN is declared if the defect diagram pixel is 1 and the reconstructed pixel is 0.
Fig. \ref{fig:pr_k_wave} shows the PR curve for each technique over all 4 tests.
Table \ref{fig:prarea} shows values of the area under the PR curves in Fig. \ref{fig:pr_k_wave}.
%
Table \ref{fig:kwavesim} shows the parameters which are used for k-wave simulation, and some of them are used as input parameters in all techniques.
Table \ref{fig:params} shows the parameters used for $l_1$-norm and MBIR in Eq. \ref{htilde}, \ref{QGGMRF}, and \ref{spatial}, and the number of iterations used.
Fig. \ref{k-wave_noisy} shows a comparison between the methods with noise added to the simulated signal of the defect diagram of Test 1 in Fig. \ref{k-wave}.

\begin{table}
	\caption{Parameter settings for k-wave simulation.}
	\label{fig:kwavesim}
	\begin{center}
		\begin{tabular}{|l|c|c|}
			\hline
			\textbf{Parameters} &\textbf{Value}&\textbf{Unit} \\ \hline
			Carrier frequency				& 52& $kHz$ \\ 
			Sampling frequency				& 1& $MHz$ \\ 
			Cement speed				& 3680&$m/s$  \\ 
			Cement density				&	1970	&	$Kg/m^3$	\\
			Cement attenuation			&	1.46e-6	&	$dB/((MHz)^y cm)$	\\
			Steel speed				& 5660&$m/s$  \\ 
			Steel density				&	8027	&	$Kg/m^3$	\\
			Steel attenuation			&	4.85e-8	&	$dB/((MHz)^y cm)$	\\
			Spatial resolution				& 1&$mm$  \\ 
			Number of columns				& 400 & - \\ 
			Number of rows					& 300& - \\ 
			Number of transducers			& 10& - \\ \hline
		\end{tabular}
	\end{center}
\end{table}
\vspace{0.1in}

\begin{table}
	\caption{Parameter settings for $l_1$-norm and MBIR reconstruct k-wave simulation data. }
	\label{fig:params}
	\begin{center}
		
		\begin{tabular}{|l|c|c|c|c|}
			\hline
			\textbf{Parameters} &\textbf{$l_1$-norm}&\textbf{MBIR}&\textbf{Unit} \\ \hline
			$\alpha_0$						&30&30& $(MHz\cdot m)^{-1}$ \\
			$p$								&-&  1.1& -\\ 
			$q$								&-&  2&-\\ 
			$T$								&-&  1&-\\ 
			$\sigma$						&0.003&  0.001&Pascal\\ 
			$c_{min}$						&-&  1&-\\ 
			$c_{max}$						&-&  10&-\\ 
			$\sigma_g$						&-&  1.3&$m^{-3}$\\ 
			$\sigma_e$						&15&  15&$m^{-3}$\\ 
			$a$								&-&  3&-\\  \hline
			
		\end{tabular}
	\end{center}
\end{table}
\vspace{0.1in}

\subsubsection{Discussion}
In Fig \ref{k-wave}, MBIR and $l_1$-norm were able to show significant enhancement over SAFT in reducing noise.
MBIR showed remarkable performance in identifying, eliminating, and distinguishing the direct arrival signal artifacts from the steel objects.
For example, in test 1, two steel plates where placed at depth 2cm.
The plates where overshadowed by the direct arrival signal artifacts in SAFT and $l_1$-norm, but appear very clearly in MBIR.
Test 2 and 3 also show similar direct arrival signal overshadowing effects for SAFT and $l_1$-norm, that are reduced for MBIR.  
In addition, the steel objects are more easily observed and recognized in $l_1$-norm and MBIR.
In Fig. \ref{fig:pr_k_wave} and Table \ref{fig:prarea}, MBIR shows better performance in the detection test with the highest PR area.

Notice that in test 4, none of the techniques were able to show the complete structure of the steel object.
They were able to show only one side of it.
This is because all three reconstruction methods reconstruct the reflections caused by discontinuous boundaries rather than recovering the actual material property at each voxel location.  

Fig. \ref{k-wave_noisy} shows the reconstruction of test 1 in Fig. \ref{k-wave} with varying signal-to-noise ratio (SNR). 
The SNR is defined as
$$
SNR = \frac{\| y \|^2}{\| w \|^2},
$$
where $y$ is the noiseless simulated output from k-wave, and $w$ is the added noise to $y$.
As the SNR decreases, the reconstruction becomes noisier for all techniques. 
However, the results show better performance in MBIR than the other techniques in reducing noise and artifacts.
\subsection{MIRA Experimental Results}
\label{results}
\begin{figure}
	\centering
	\includegraphics[width=0.3\textwidth]{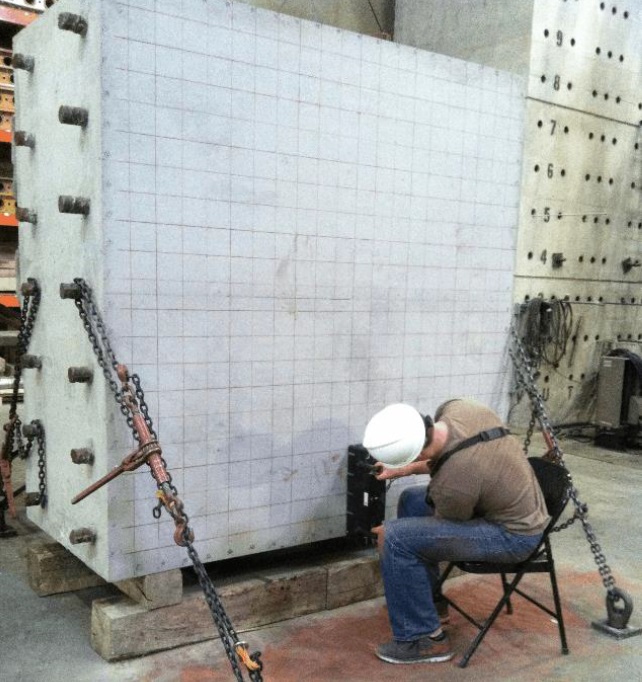}
	\caption{The concrete specimen used for the experimental data \cite{clayton_barker_santos-villalobos_albright_hoegh_khazanovich_2015}. 20 defects are embedded in the specimen.}
	\label{fig:concrete_4}
\end{figure}

\begin{figure}
	\centering
	\includegraphics[width=0.3\textwidth]{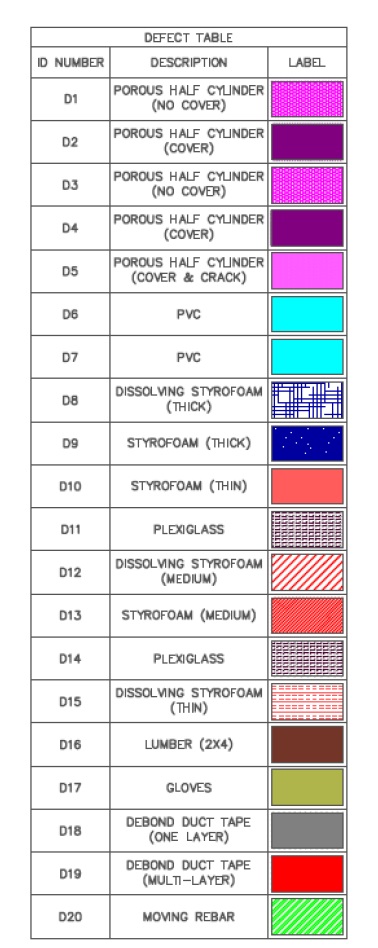}
	\caption{Type and legend for each defect \cite{clayton_barker_santos-villalobos_albright_hoegh_khazanovich_2015}. These defects are embedded in the concrete specimen.}
	\label{fig:defect}
\end{figure}

\begin{figure}
	\centering
	\includegraphics[width=0.3\textwidth]{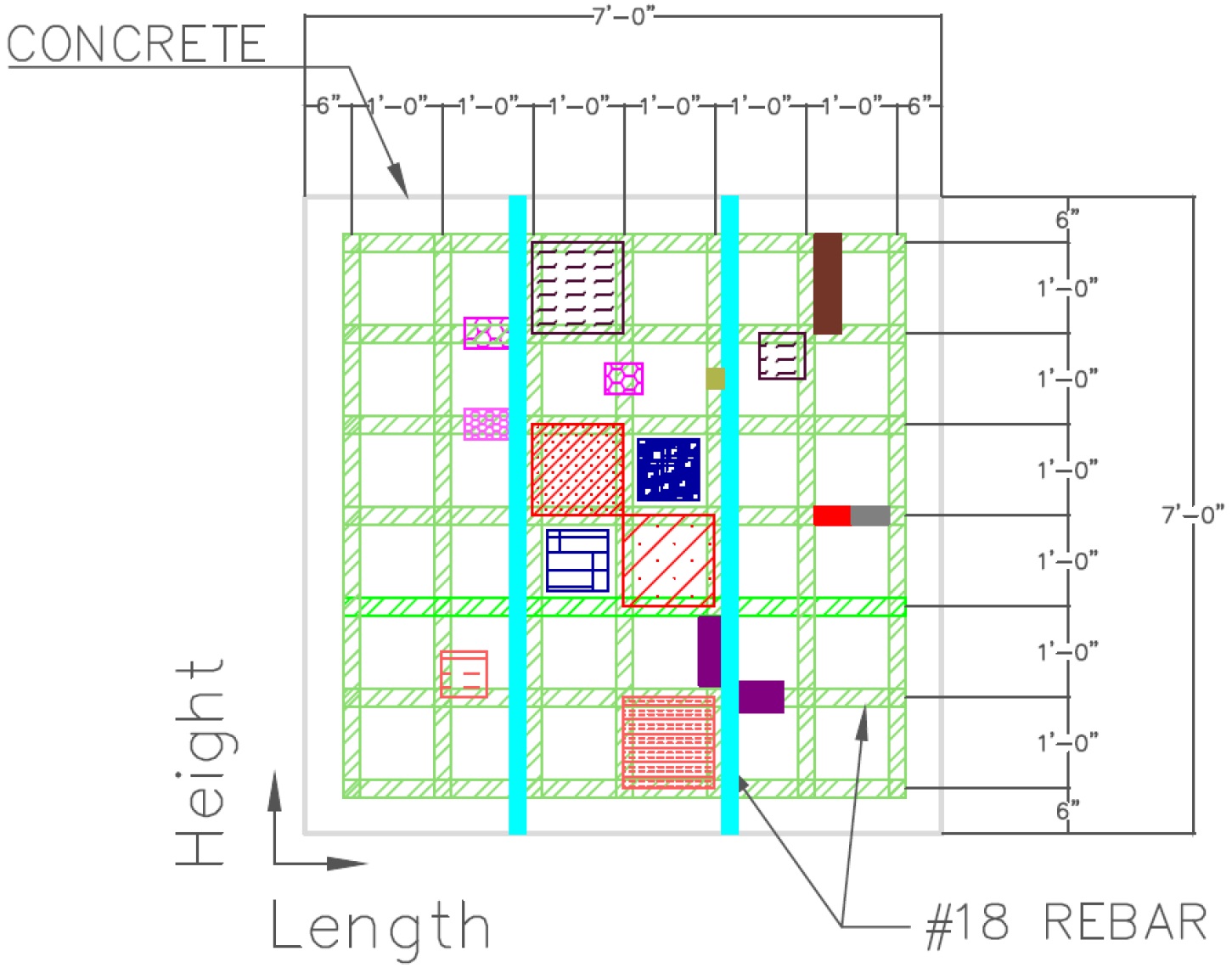}
	\caption{Smooth side view of defects \cite{clayton_barker_santos-villalobos_albright_hoegh_khazanovich_2015}. The location of the defects is approximated due to possible displacement while pouring the cement.}
	\label{fig:defect3}
\end{figure}

\begin{figure}
	\centering
	\includegraphics[width=0.3\textwidth]{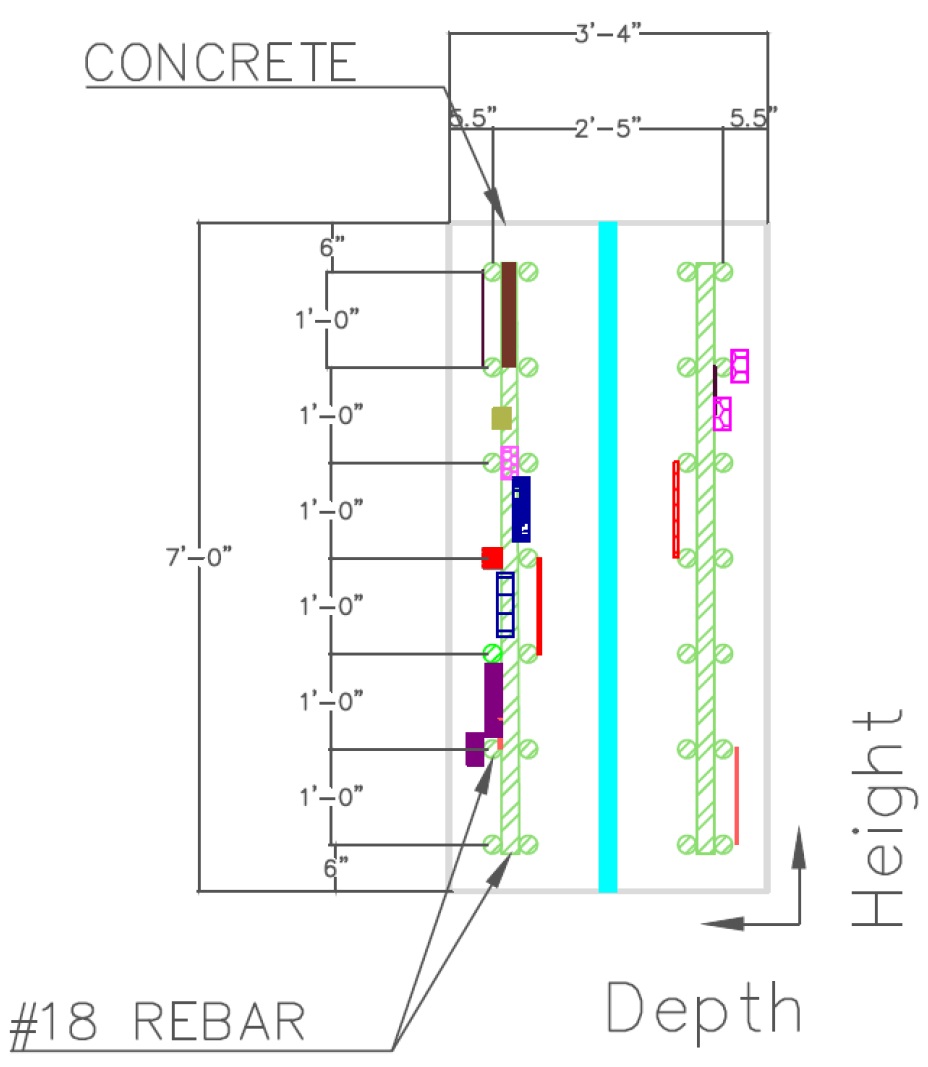}
	\caption{Depth view of defects, smooth side on the right and rough side on the left, \cite{clayton_barker_santos-villalobos_albright_hoegh_khazanovich_2015}. The location of the defects is approximated due to possible displacement while pouring the cement.}
	\label{fig:defect4}
\end{figure}

\begin{figure}
	\centering
	\includegraphics[width=0.3\textwidth]{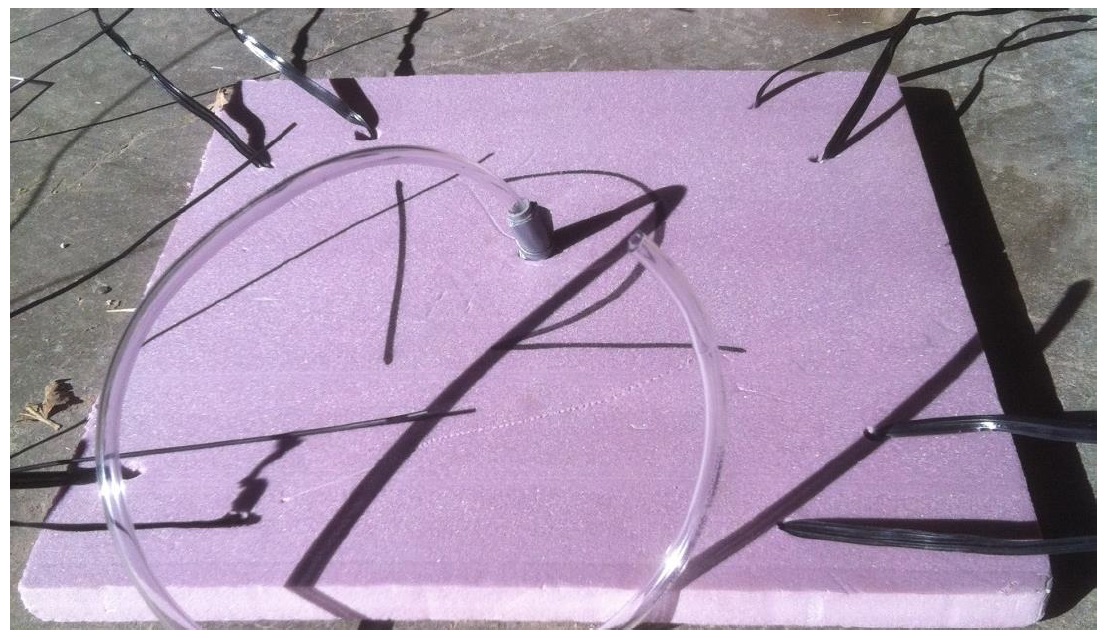}
	\caption{A picture of defect 12 before embedding it in the specimen, \cite{clayton_barker_santos-villalobos_albright_hoegh_khazanovich_2015}. It is made of dissolving styrofoam.}
	\label{fig:defect5}
\end{figure}

Experimental results have been obtained from a designed thick concrete specimen \cite{clayton_barker_santos-villalobos_albright_hoegh_khazanovich_2015}. 
The height and width of the specimen is 84 inches, Fig. \ref{fig:concrete_4}. The depth of the specimen is 40 inches. 
Each side of the block is gridded with 4-inch squares producing 21 rows and columns.
The specimen has been heavily reinforced with steel rebars horizontally and vertically with 1 ft separation in both sides. 
One side is ``smooth'' and the other is ``rough'' which refer to the physical characteristic of the concrete surface due to pouring.
Also, Fig. \ref{fig:defect3} and Fig. \ref{fig:defect4} show diagrams of the steel rebars in green color with more details.
The specimen has been embedded with designed defects placed in specific locations. 
The type and location of the defects are shown in Fig. \ref{fig:defect}, \ref{fig:defect3}, \ref{fig:defect4}, and \ref{fig:defect5} \cite{clayton_barker_santos-villalobos_albright_hoegh_khazanovich_2015}.
The specified location of the defects might be different from the real location due to possible displacement while pouring the cement.

The defects are designed to simulate real defects that can occur due to construction process, cumulative deterioration, or degradation of concrete.
Both sides are scanned horizontally and vertically.
There are four types of scanning modes: smooth-horizontal (SH), smooth-vertical (SV), rough-horizontal (RH), and rough-vertical (RV).
Each mode divides the whole side into 19 sets which adds up to 76 sets.
However, only 73 sets are used and are arranged in this order: sets 1 to 18 for SH, sets 19 to 37 for SV, sets 38 to 56 for RH, and sets 57 to 73 for RV.
Each set scans the side from left to right or from bottom to top, depending on the orientation of the mode, with 18 positions.
The first and last positions are centered at 8 inches from the edge.
The rest of the positions are centered with a 4-inch shift from the previous position, hence the 18 positions.

\begin{figure}
	\centering
	\includegraphics[width=0.4\textwidth]{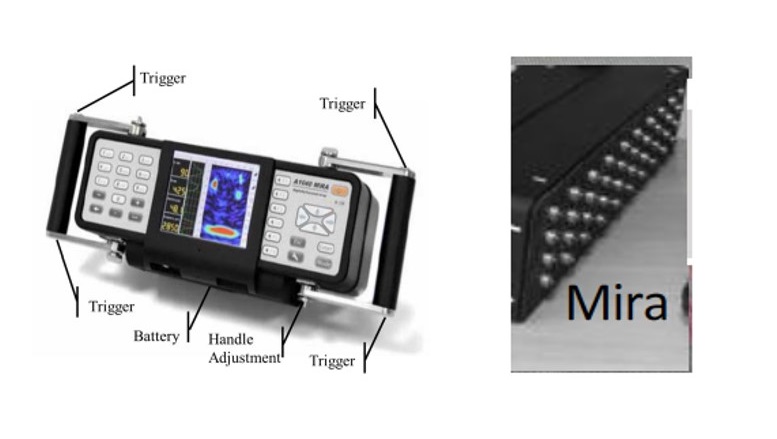}
	\caption{A picture of the MIRA device used for the experimental data. The device has 10 columns of transducers, where each column acts as a single transducer.}
	\label{MIRA}
\end{figure}

The MIRA system has been used to collect the data, Fig. \ref{MIRA}.
The MIRA device contains 10 columns or channels separated by 40 mm where each channel contains 4 dry contact points with 2 mm radius that acts as transmitters or receivers.
Only distinct pairs, 45 pairs, are used in the reconstruction results for all techniques.
Each position produces an image of width 40 cm and depth 120 cm with 1 cm resolution.
SAFT requires approximately 0.03 seconds to reconstruct an individual image, and MBIR requires approximately 5 seconds for the same image.
There are four different techniques used to reconstruct the data: SAFT, $l_1$-norm, 2D MBIR, and 2.5D MBIR.
For SAFT and $l_1$-norm, all images of the set are stitched together to produce the complete cross section of the set with 210cm-width inches and 120cm-depth.
For 2D MBIR, and 2.5D MBIR, the joint-MAP stitching is used to reconstruct the whole cross section at once instead of regular stitching.
The joint-MAP stitching reduced the MBIR processing time of the whole data from about 110 minutes to about 87 minutes (about $30\%$ lower).
However, the 2.5D MBIR increased the processing time from 87 minutes to 126 minutes (about $45\%$ higher).
All the techniques were implemented in Windows with a 6th generation Intel core i7-6500U processor with 4 MB cache, 2/4 core/threads, and 2.5 GHz CPU.
For both MBIR results, $\sigma$ in Eq. \ref{map} is estimated  as described in \cite{BoumanBook2013}. 
Note that the measurements and reconstruction use the metric system while the design of the specimen uses the imperial system.
The transmitter emits a signal with carrier frequency of 52 kHz, and the sampling frequency of the receiver is 1 MHz.
The acoustic speed is assumed to be $2620 \frac{m}{s}$.
Each distinct pair produces 2048 samples of data where the first 27 samples are ignored due to trigger synchronization.
The data is, then, down-sampled to 200 kHz and 409 samples and reconstructed using all techniques.

Fig. \ref{quali} shows the reconstruction results. 
The reconstruction 2D voxel spacing is 1 cm for all techniques.
The rows are ordered from top to bottom.
The first row shows the defect diagram and the position of the defects. 
The scanning of the cross sections was performed at the top of the image from left to right. 
The second row is the instantaneous envelope of SAFT reconstruction. 
The third row is $l_1$-norm reconstruction. 
The fourth row is 2D MBIR reconstruction. 
The fifth row is 2.5D MBIR reconstruction. 
Note that the defect diagram shows the steel rebars as dotted circles or dotted rectangles.
The steel rebars might appear in all reconstructions as small horizontal dots or a horizontal line at the top, but the bottom steel rebars barely appear in all techniques due to their weak reflection.
Table \ref{fig:MIRA} shows the common parameter settings for all techniques.
Table \ref{fig:MIRAMBIR} shows the $l_1$-norm and MBIR parameter settings for Eq. \ref{htilde}, \ref{QGGMRF}, and \ref{spatial}, $\gamma$ in section \ref{prior_model}, and the number of iterations used.

\begin{figure}
	\centering
	\includegraphics[width=0.4\textwidth]{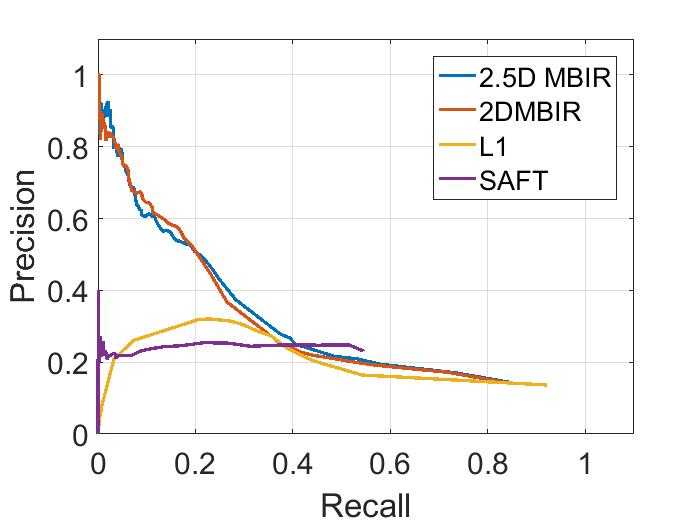}
	\caption{PR curves for each technique over all 73 experimental data sets. 2.5D and 2D MBIR outperforms the other techniques.}
	\label{fig:miss_avg}
\end{figure}

Because the position of the targets in the defect diagram is not precise,
the detection test was done using a component wise approach rather than the pixel-wise approach used for the k-wave data.
Each image is segmented into connected components using the standard Matlab functions ``edge'' and ``imfill''.
Next, the maximum value and weighted centroid for each connected component is stored.
Next, a search is performed pairing targets from the defect diagram to connected components from the reconstruction in the following way:
A connected component is mapped to a particular target if its centroid is both the closest among all detected components to the target's centroid, and it is within 10 cm of the target's centroid.
The pairing will help us later in deciding if the detection is TP or FP.
Next, for each technique, all the images are normalized by dividing them with the maximum value of them all.
Thresholds from 1 to 0 with step 0.001 are applied to all images.
For each threshold, a TP is declared if the maximum value of a paired connected component is equal or greater than the threshold.
A FP is declared if the maximum value of an unpaired connected component is equal or greater than the threshold.
The FN is calculated by subtracting the number of TP's from the number of targets.
Fig. \ref{fig:miss_avg} shows the PR curve for each technique over all 73 experimental data sets.
Table \ref{fig:prarea} shows value of the area under the PR curves in Fig. \ref{fig:miss_avg}.
For 2D MBIR, the value of $\sigma_g$ that maximized the PR area was chosen.
For $l_1$-norm, the value of $\sigma_e$ that maximized the PR area was chosen.
For 2.5D MBIR, the values of $\sigma_g$ and $\gamma$ that maximized the PR area were chosen.

\subsubsection{Discussion}
\begin{figure*}
	\begin{center}
		\begin{tabular}{@{}c@{}c@{}c@{}c@{}c@{}}
			&smooth-hor-set8&smooth-ver-set9&rough-hor-set8&rough-ver-set7
			\tabularnewline
			\rot{Defect Diagram}&
			\includegraphics[align = c,width=0.2\textwidth]{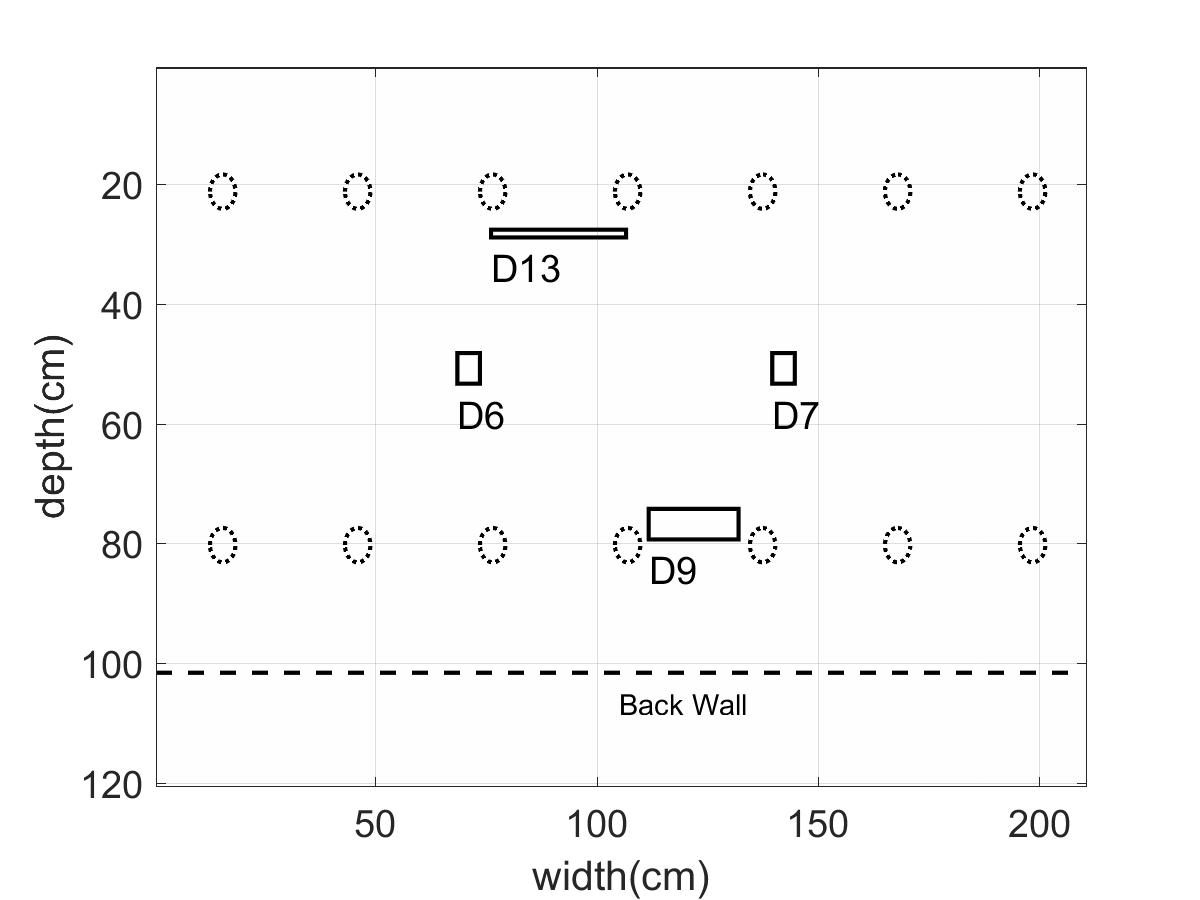}&
			\includegraphics[align = c,width=0.2\textwidth]{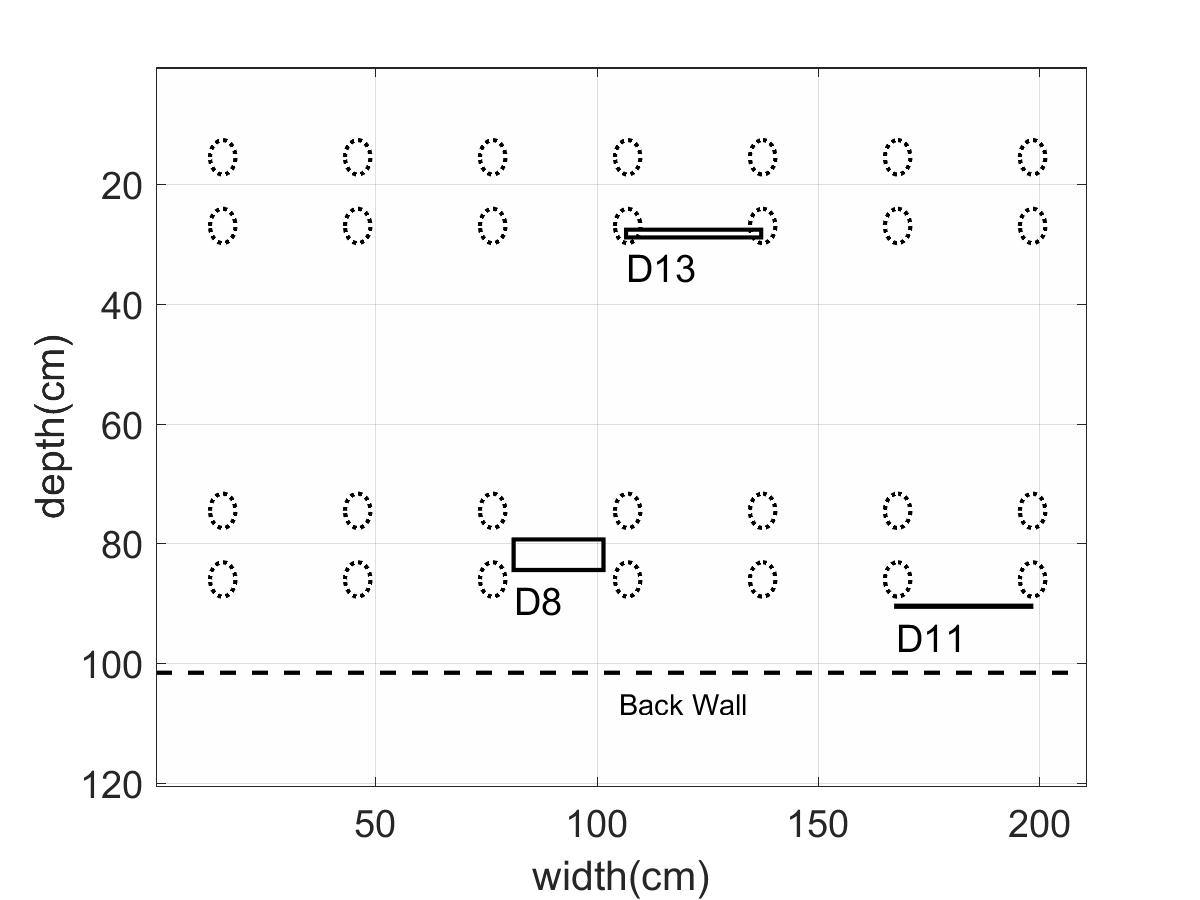}&
			\includegraphics[align = c,width=0.2\textwidth]{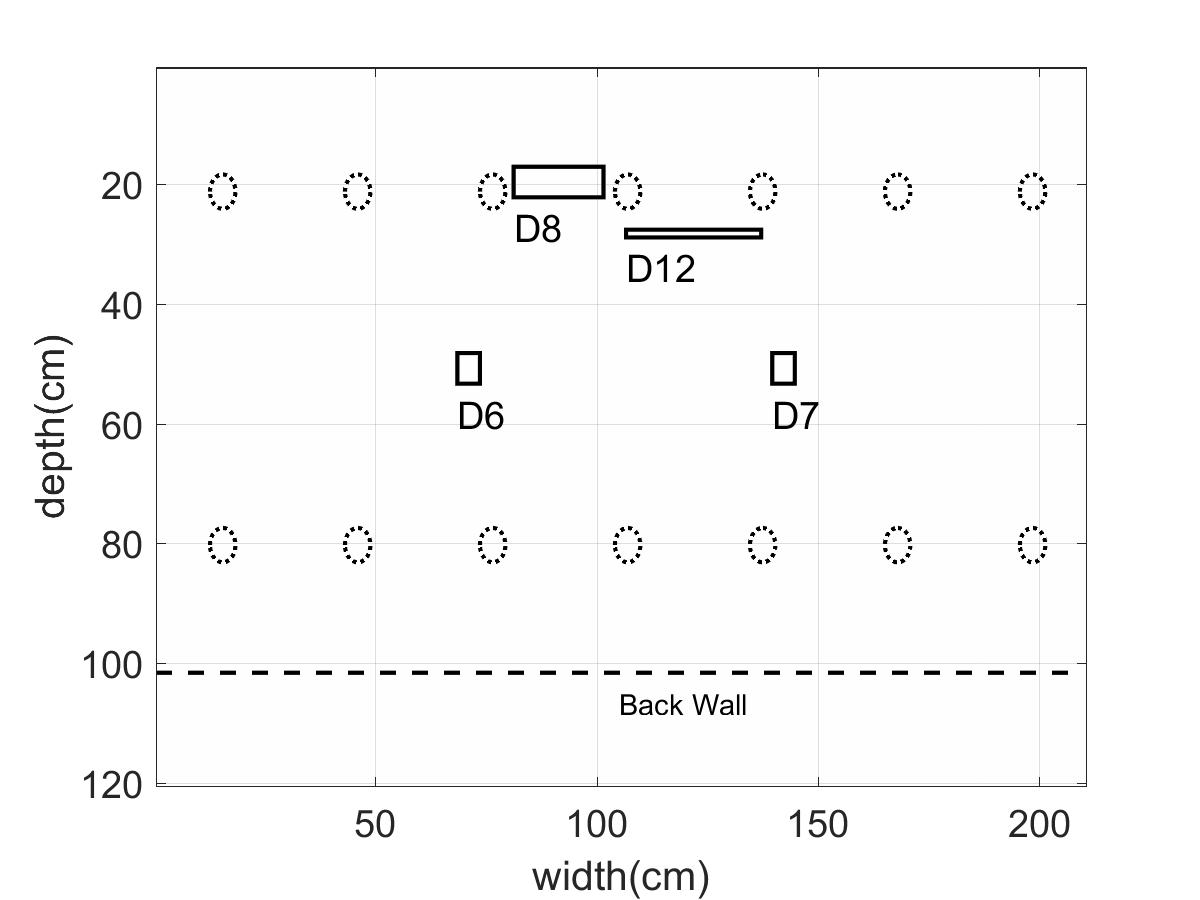}&
			\includegraphics[align = c,width=0.2\textwidth]{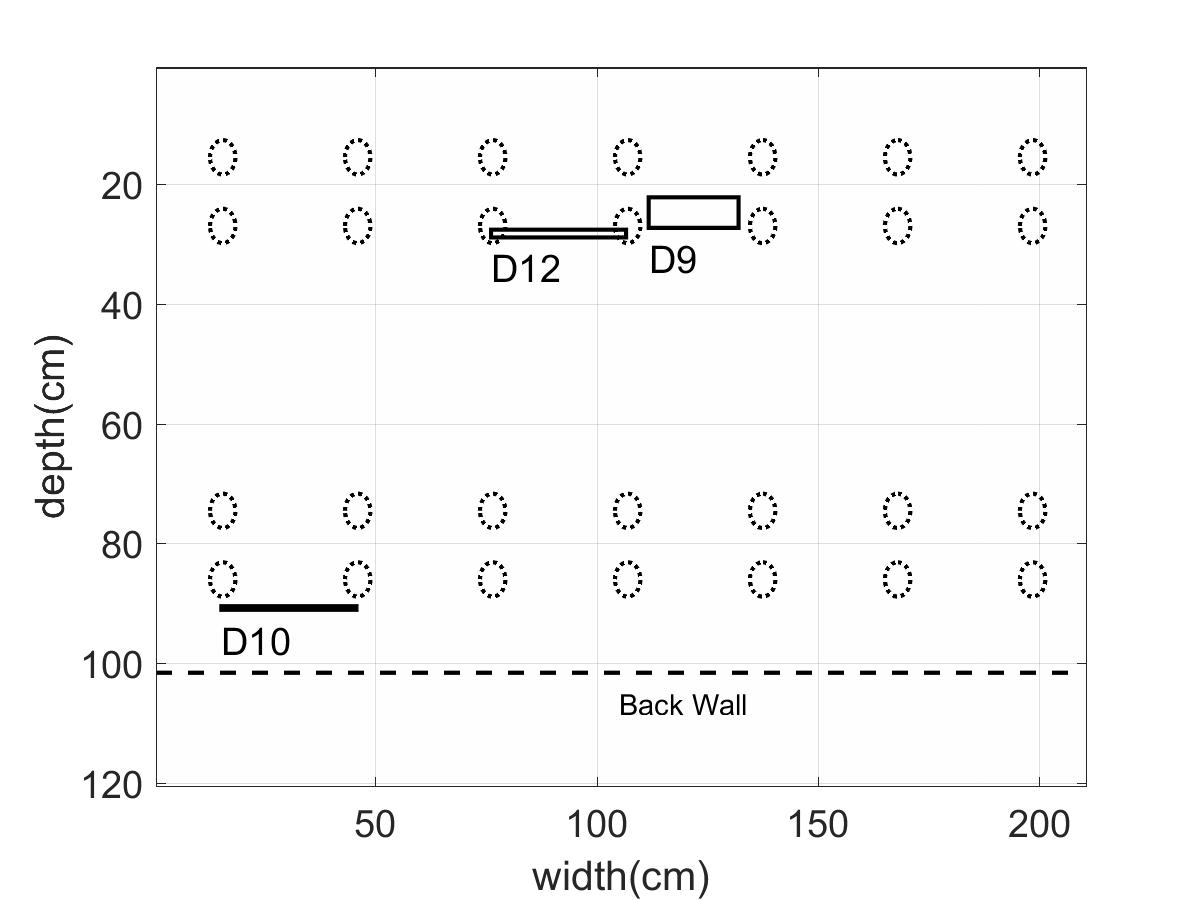}
			\tabularnewline

			\rot{SAFT} &
			\includegraphics[align = c,width=0.2\textwidth]{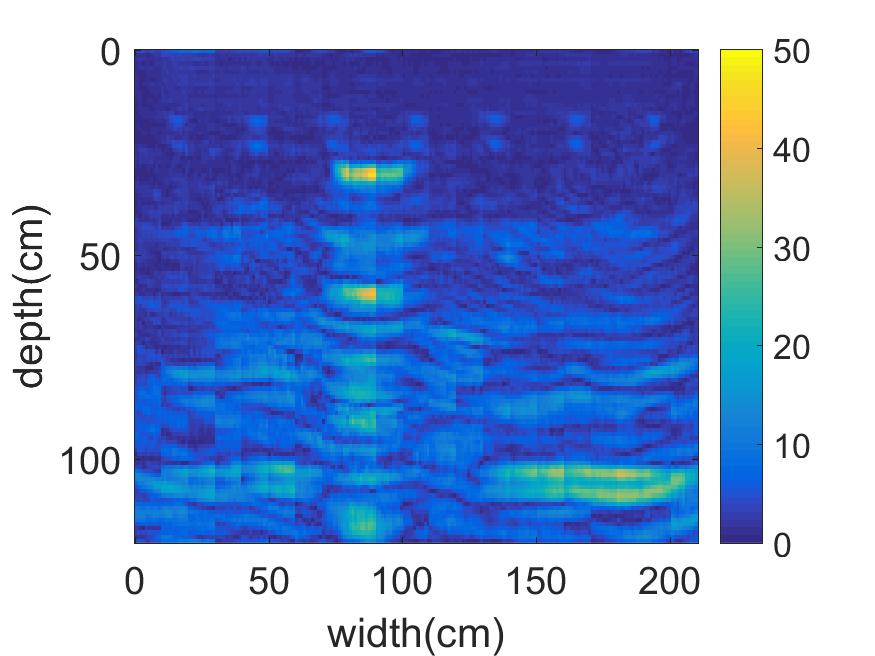}&
			\includegraphics[align = c,width=0.2\textwidth]{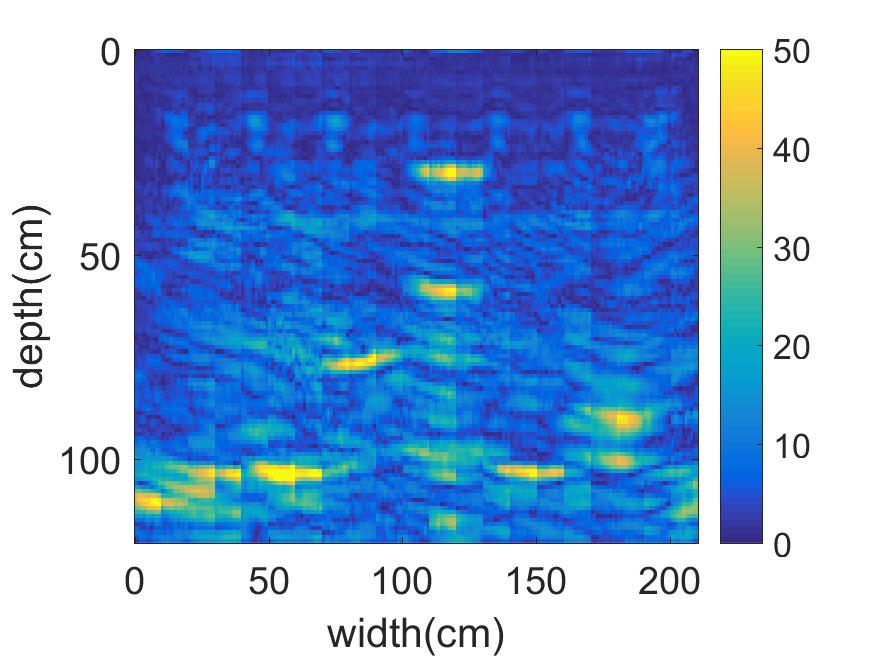}&
			\includegraphics[align = c,width=0.2\textwidth]{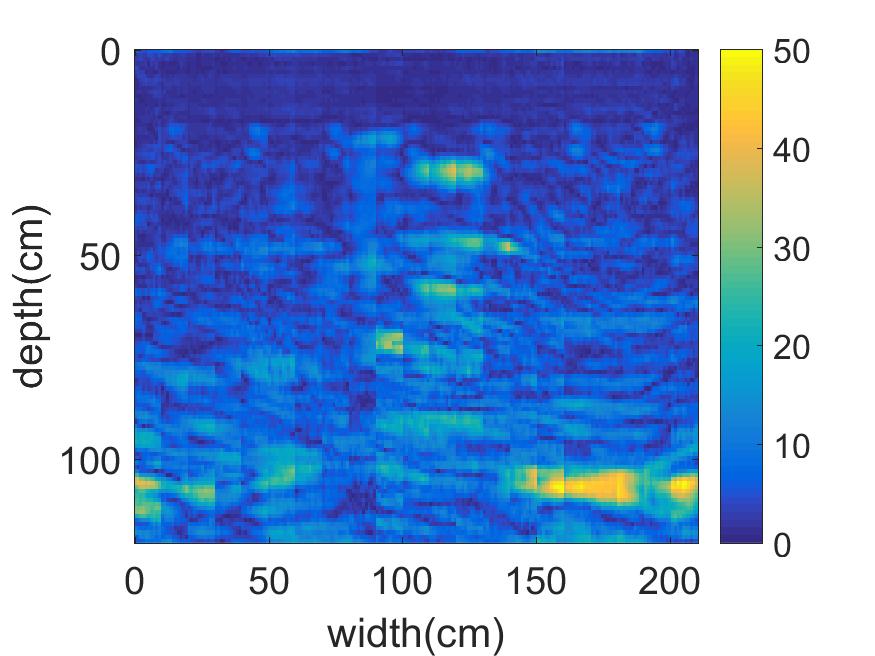}&
			\includegraphics[align = c,width=0.2\textwidth]{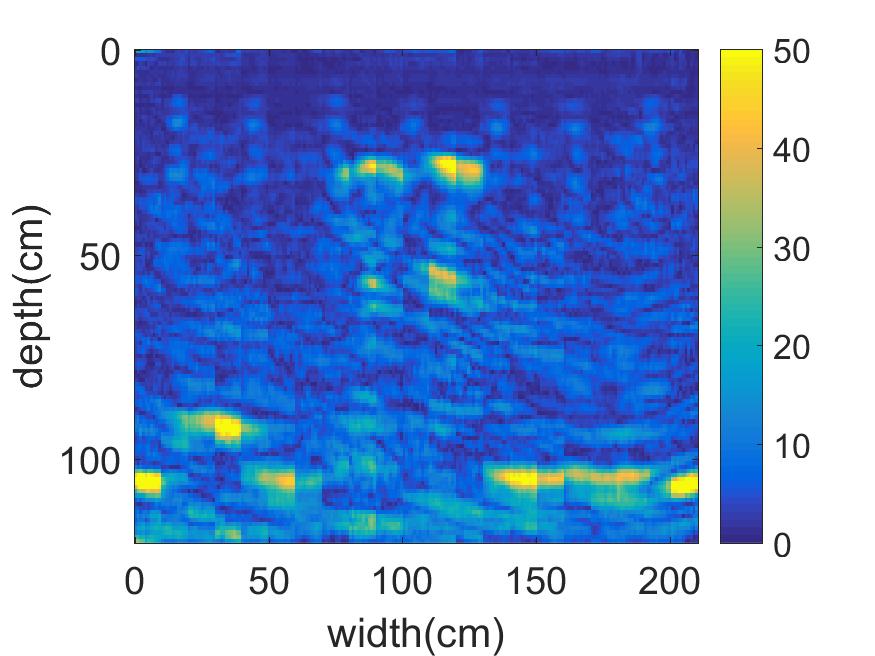}
			\tabularnewline

			\rot{L1 Norm} &
			\includegraphics[align = c,width=0.2\textwidth]{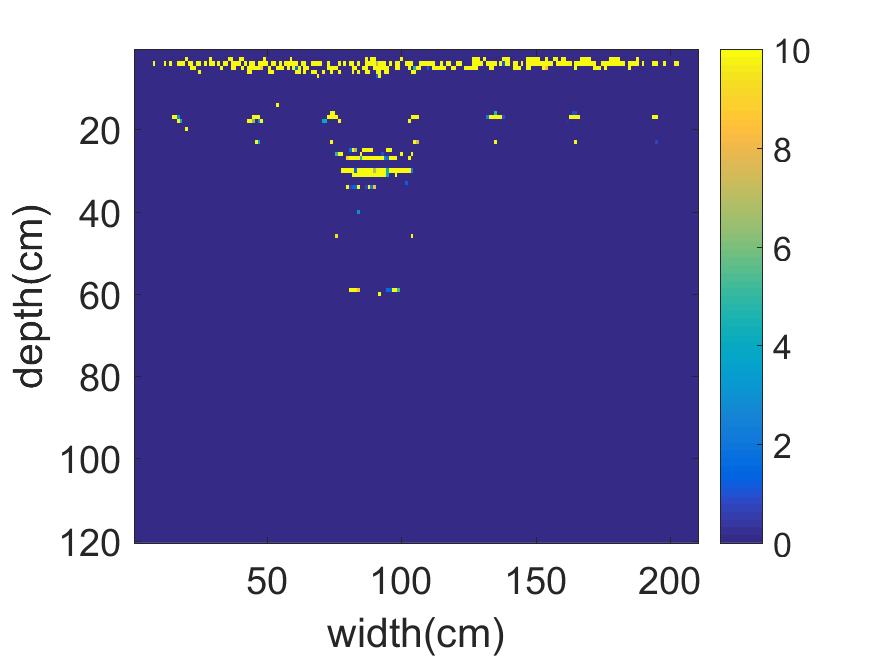}&
			\includegraphics[align = c,width=0.2\textwidth]{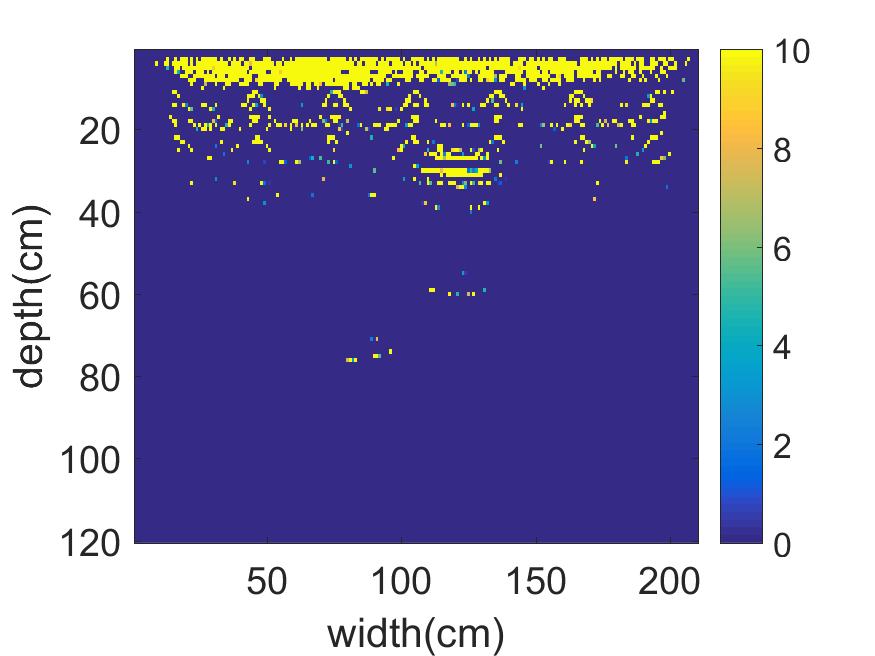}&
			\includegraphics[align = c,width=0.2\textwidth]{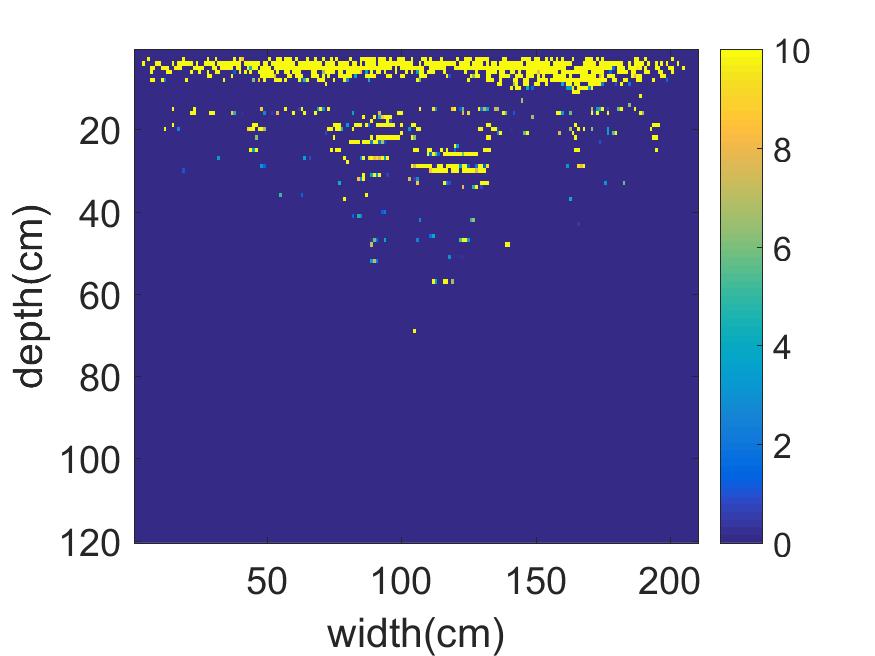}&
			\includegraphics[align = c,width=0.2\textwidth]{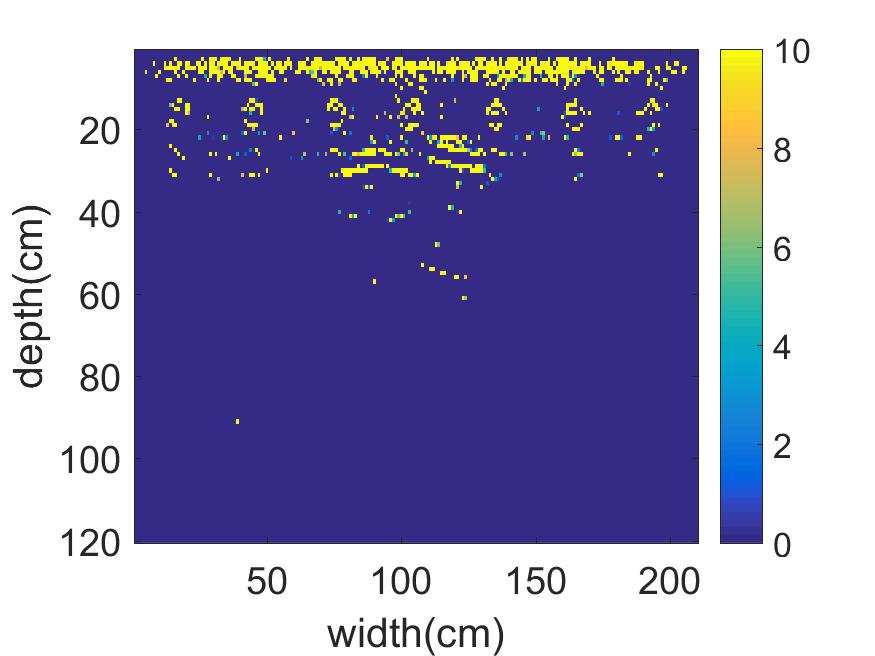}
			\tabularnewline
			
			\rot{2D MBIR} &
			\includegraphics[align = c,width=0.2\textwidth]{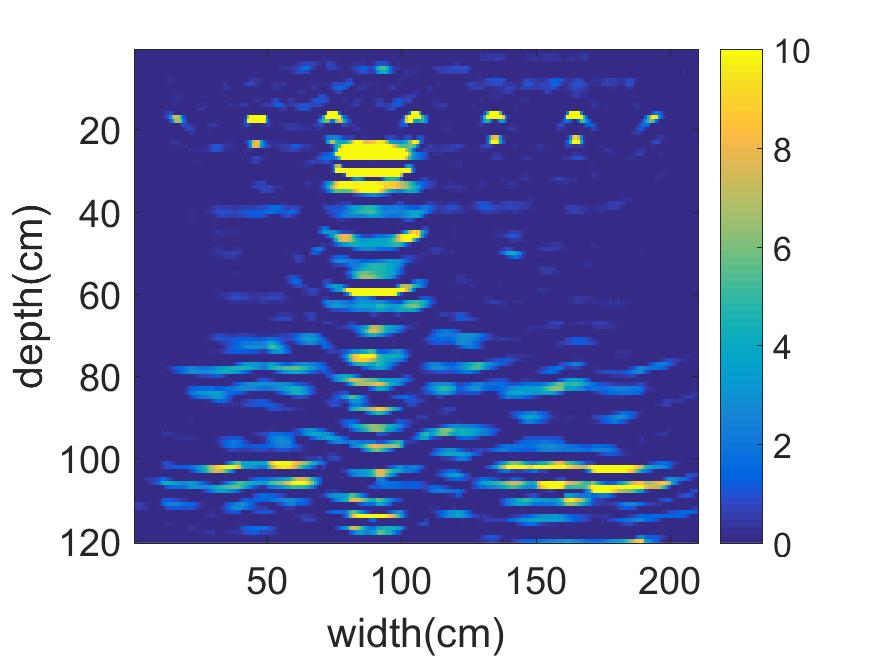}&
			\includegraphics[align = c,width=0.2\textwidth]{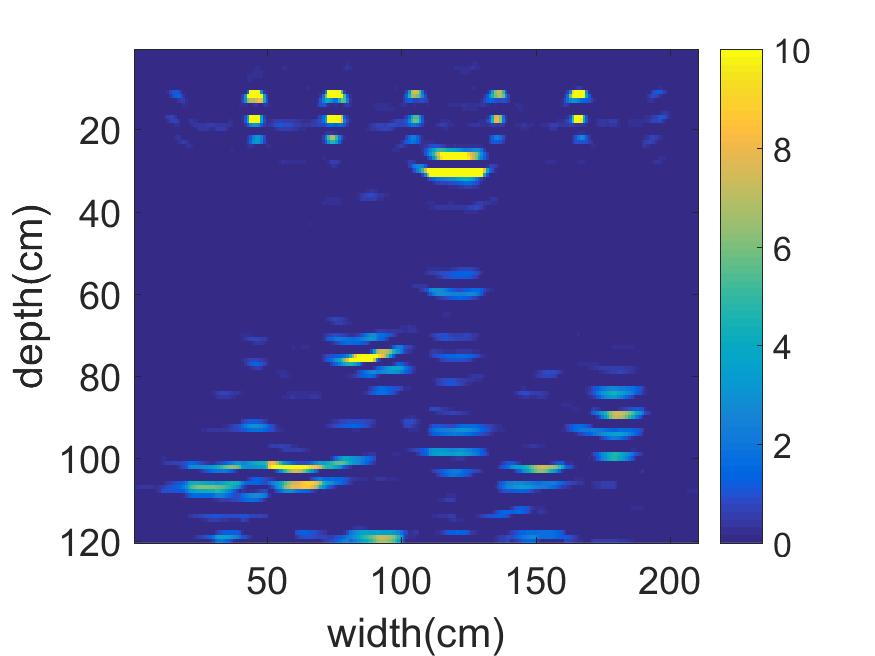}&
			\includegraphics[align = c,width=0.2\textwidth]{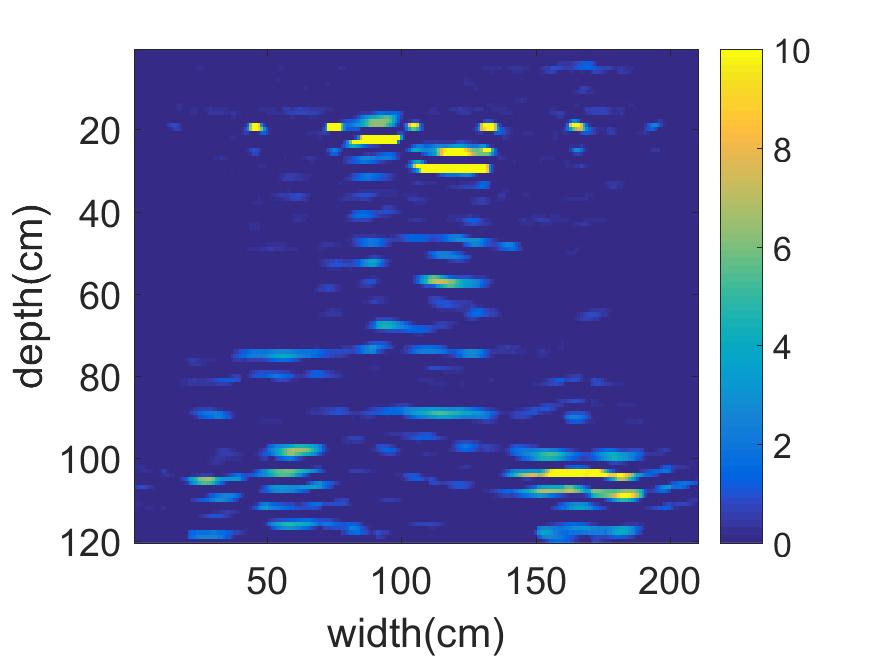}&
			\includegraphics[align = c,width=0.2\textwidth]{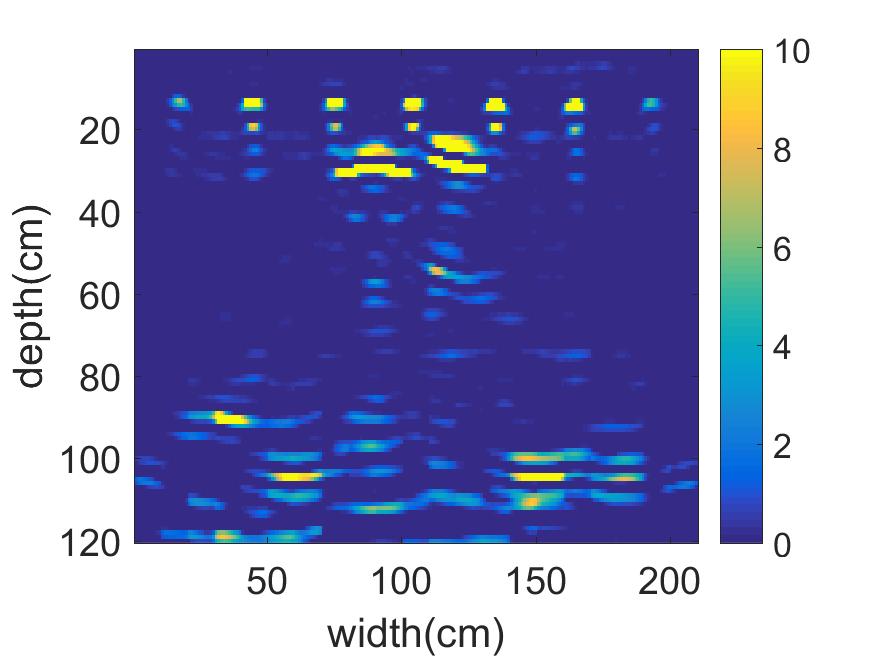}
			\tabularnewline
			\rot{2.5D MBIR} &
			\includegraphics[align = c,width=0.2\textwidth]{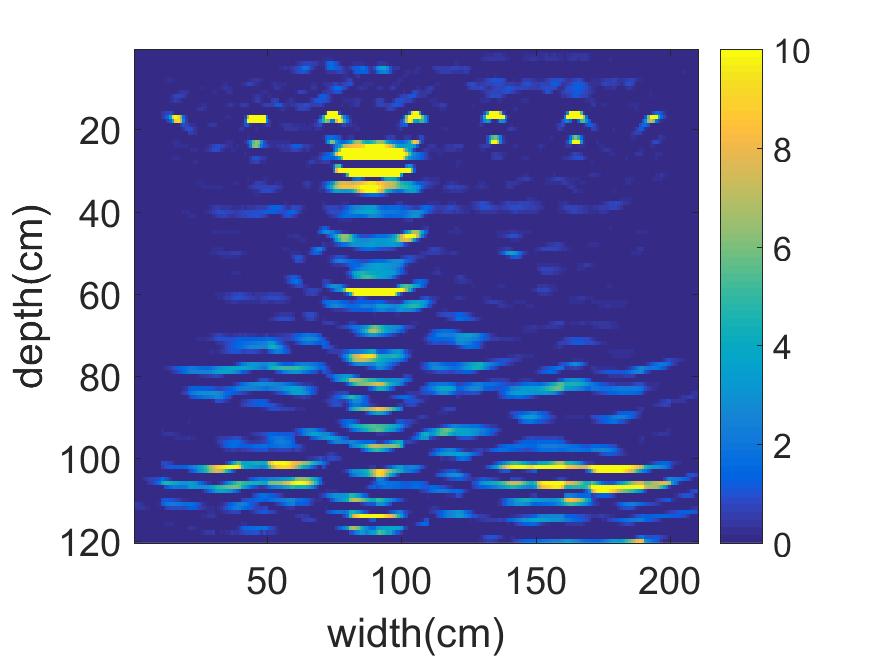}&
			\includegraphics[align = c,width=0.2\textwidth]{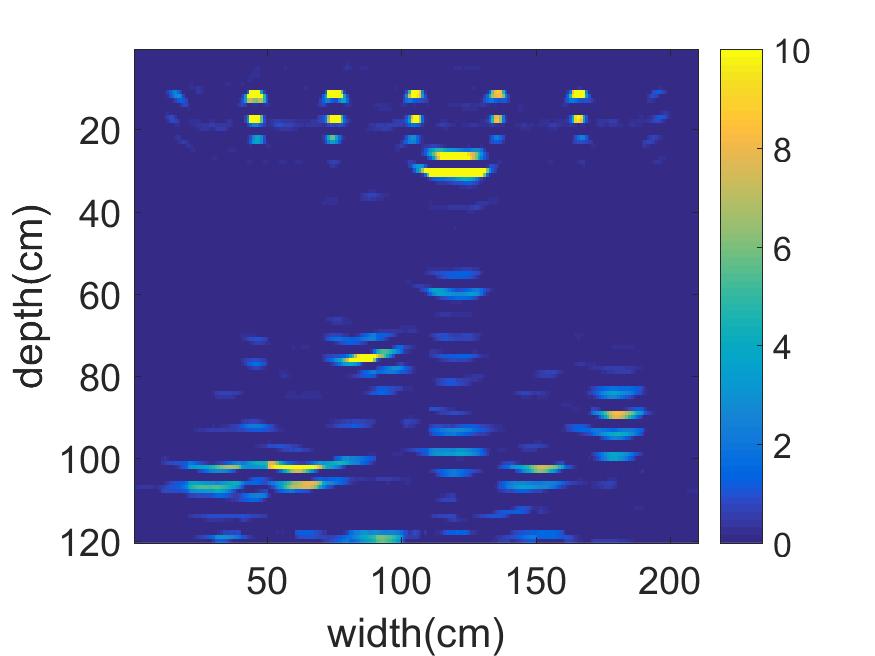}&
			\includegraphics[align = c,width=0.2\textwidth]{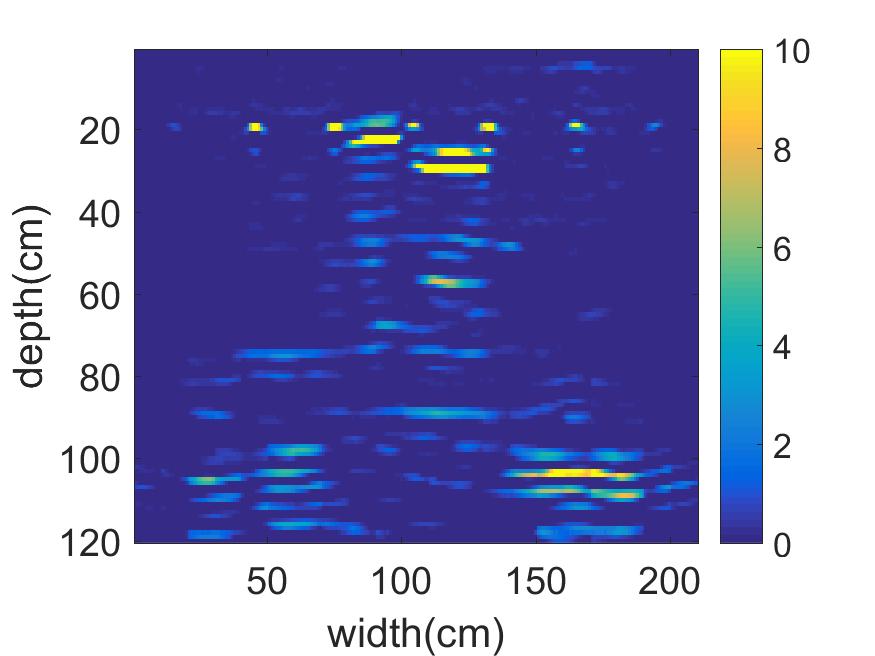}&
			\includegraphics[align = c,width=0.2\textwidth]{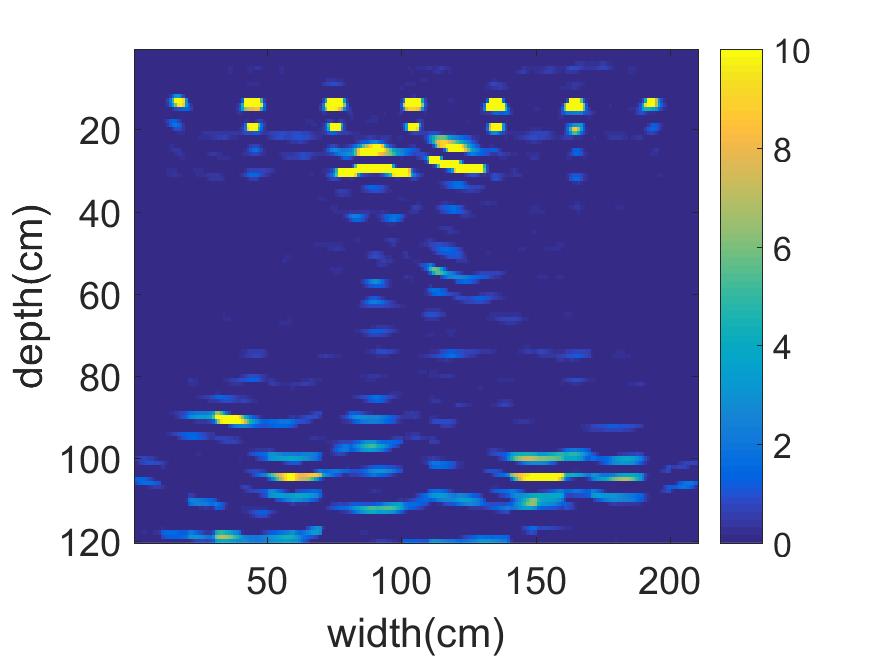}
			\tabularnewline

		\end{tabular}
	\end{center}
	\vspace{-0.15in}
	\caption{\label{quali}
		\scriptsize
		{Comparison between all reconstruction results from the MIRA experimental data: the first row from the top is the position of the defects, the second row is SAFT reconstruction, the third row is $l_1$-norm reconstruction, the fourth row is 2D MBIR reconstruction, and the fifth row is 2.5D MBIR reconstruction. 2.5D and 2D MBIR tend to produce results with less noise and artifacts compared to other techniques. }}
\end{figure*}

In Fig. \ref{quali}, MBIR shows great enhancement in reducing artifacts and reducing noise compared with SAFT and $l_1$-norm.
SAFT and MBIR techniques were able to show the back wall of the specimen.
The back wall is located at depth 100 cm.
The detection test showed great performance of 2.5D and 2D MBIR over 
all techniques with 2.5D MBIR  being slightly better than 2D MBIR.

Since all three algorithms are based on a linear forward model, they all exhibit certain reconstruction artifacts such as multiple reflection echos of a single defect. 
For example, multiple echos appeared of defect 13 in SV8 for all techniques.


\begin{table}
	\caption{Parameter settings used for all techniques to reconstruct the experimental MIRA data.}
	\label{fig:MIRA}
	\begin{center}
		\begin{tabular}{|l|c|c|}
			\hline
			\textbf{Parameters} &\textbf{Value}&\textbf{Unit} \\ \hline
			Carrier frequency				& 52& $kHz$ \\ 
			Sampling frequency				& 200& $kHz$ \\ 
			Cement p-wave speed				& 2620&$m/s$  \\ 
			Reconstruction resolution		& 1&$cm$  \\ 
			Number of columns				& 210 & - \\ 
			Number of rows					& 120& - \\ \hline
		\end{tabular}
	\end{center}
\end{table}
\vspace{0.1in}

\begin{table*}
	\caption{The $l_1$-norm, 2D MBIR, and 2.5D MBIR parameters settings used to reconstruct the experimental MIRA data.}
	\label{fig:MIRAMBIR}
	\begin{center}
		
		\begin{tabular}{|l|c|c|c|c|c|}
			\hline
			\textbf{Parameters} &\textbf{$l_1$-norm}&\textbf{2D MBIR}&\textbf{2.5D MBIR}&\textbf{Unit} \\ \hline
			$\alpha_0$						&30&30&30& $(MHz\cdot m)^{-1}$ \\
			$p$								&-&1.1&  1.1& -\\ 
			$q$								&-&2&  2&-\\ 
			$T$								&-&1&  1&-\\ 
			$\sigma$						&1&Estimated&  Estimated&Pascal\\ 
			$c_{min}$					&-&1&  1&-\\ 
			$c_{max}$					&-&10&  10&-\\ 
			$\sigma_g$						&-&3&  3&$m^{-3}$\\ 
			$\sigma_e$						&15&15&  15&$m^{-3}$\\ 
			$a$								&-&3&  3&-\\ 
			$\gamma$						&-&0&  0.5&-\\ \hline
			
		\end{tabular}
	\end{center}
\end{table*}
\vspace{0.1in}

\begin{table}
	\caption{Precision vs recall area for all techniques in Fig. \ref{fig:pr_k_wave} and Fig. \ref{fig:miss_avg}. MBIR has the highest PR area.}
	\label{fig:prarea}
	\begin{center}
		\begin{tabular}{|c|c|c|c|c|}
			\hline
			 &SAFT&$l_1$-norm&2D MBIR&2.5D MBIR\\ \hline
			PR area for k-wave data& 0.1236 &0.2131&  0.3476&- \\ \hline
			PR area for MIRA data& 0.1323 &0.1932&  0.2836&0.2908 \\ \hline
		\end{tabular}
	\end{center}
\end{table}
\begin{figure*}
	\begin{center}\footnotesize
		\begin{tabular}{@{}c@{}c@{}c@{}c@{}}
			\includegraphics[align = c,width=0.2\textwidth]{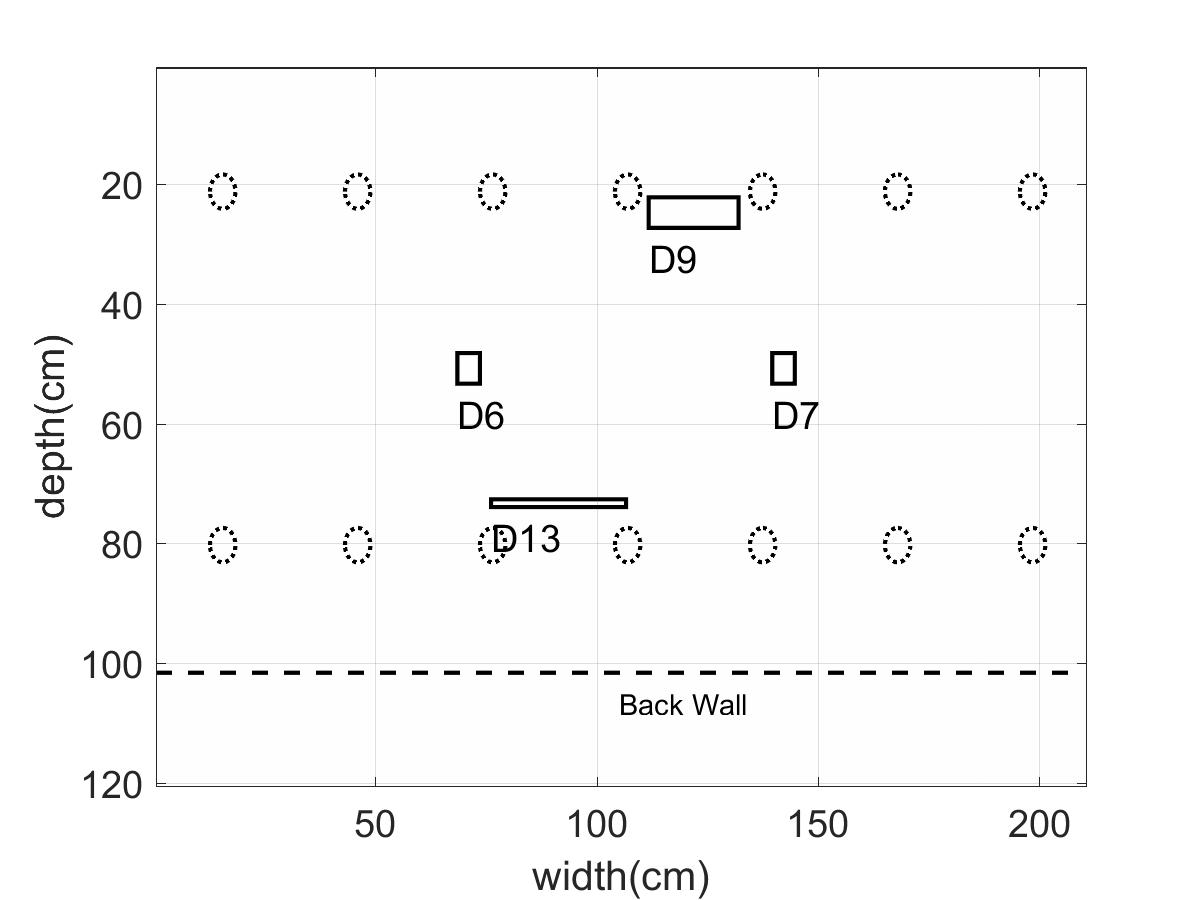} &
			\includegraphics[align = c,width=0.2\textwidth]{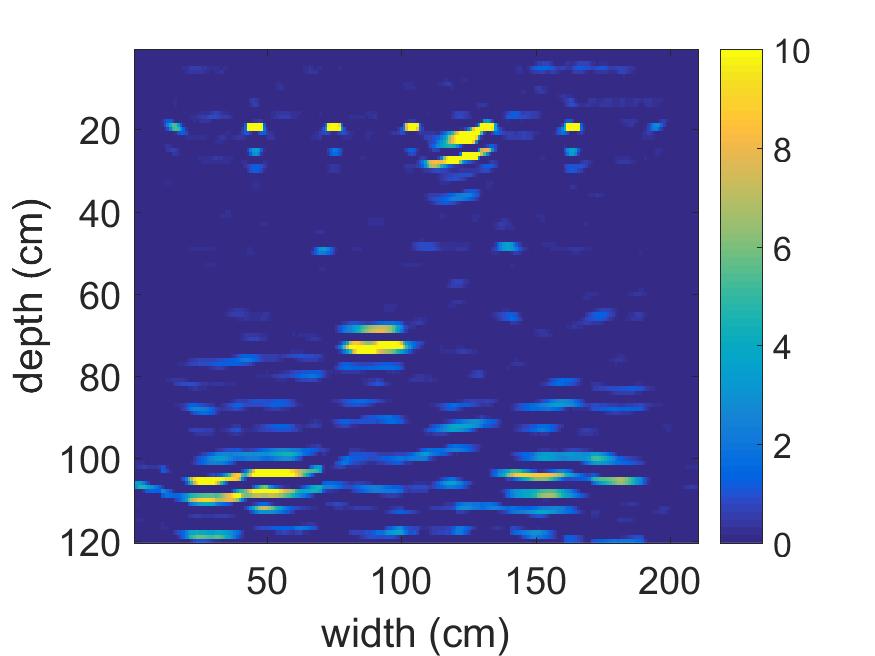} &
			\includegraphics[align = c,width=0.2\textwidth]{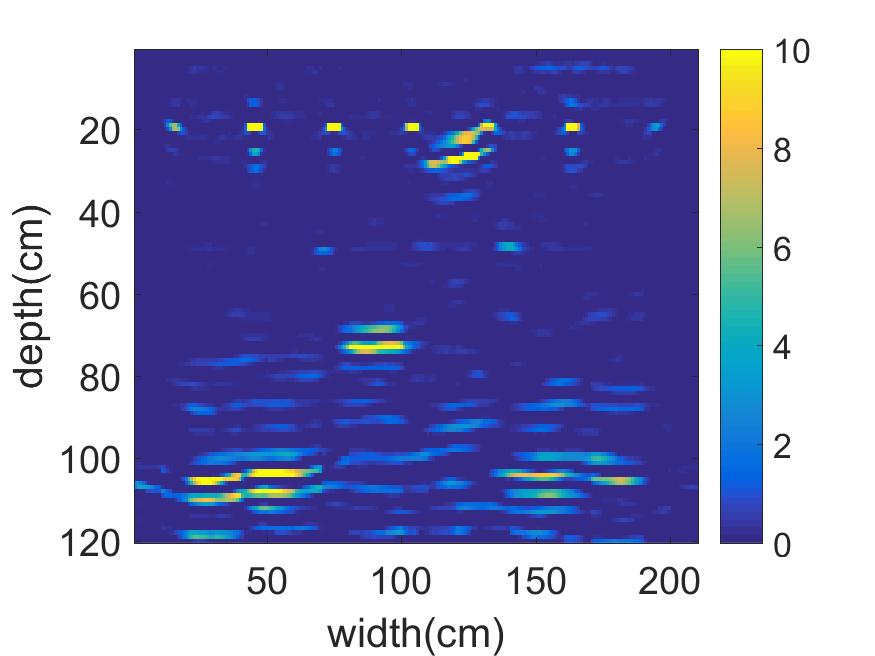} &
			\includegraphics[align = c,width=0.2\textwidth]{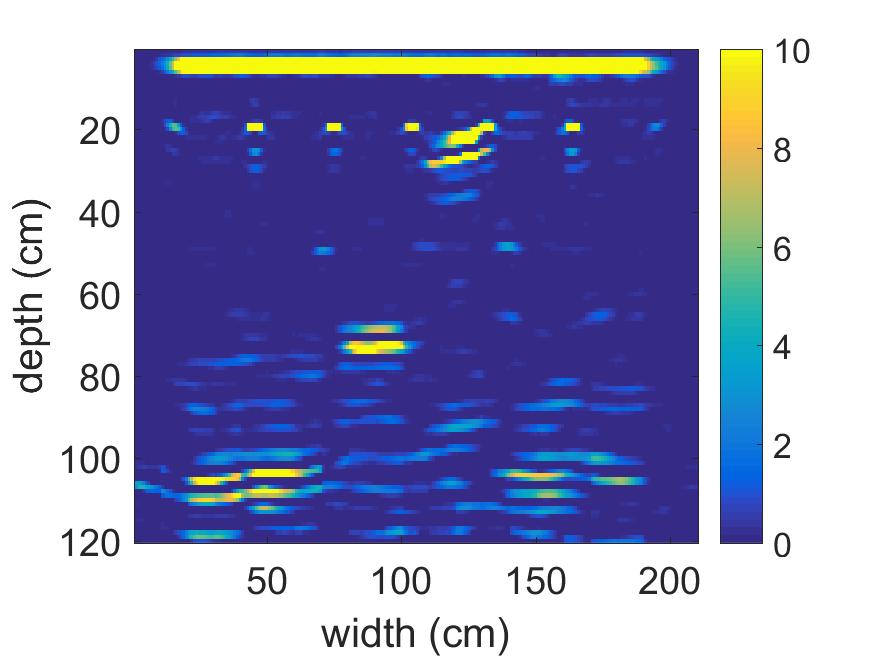}
			
			\tabularnewline
			(a) & (b) & (c) & (d)
			\tabularnewline
			\includegraphics[align = c,width=0.2\textwidth]{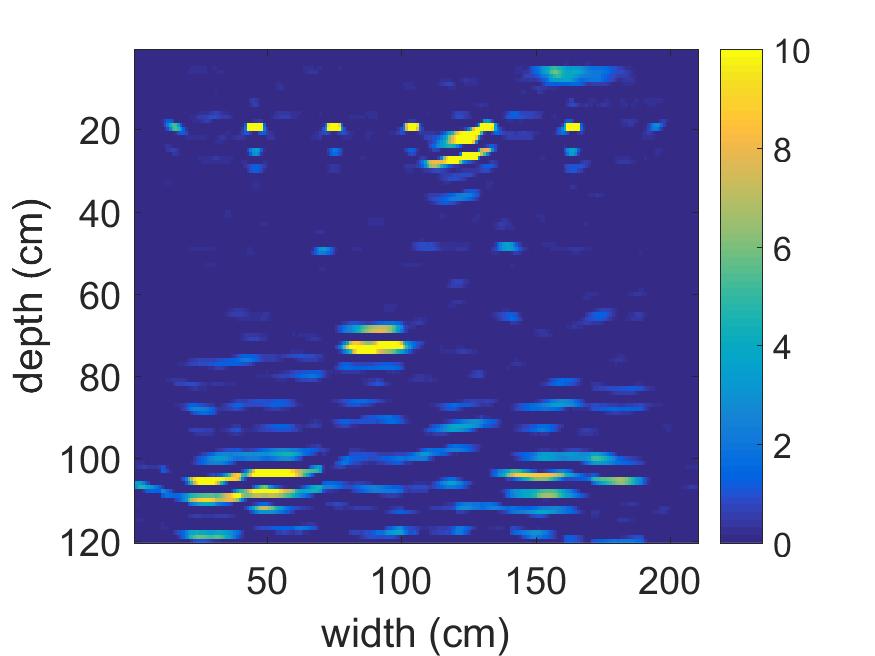} &
			\includegraphics[align = c,width=0.2\textwidth]{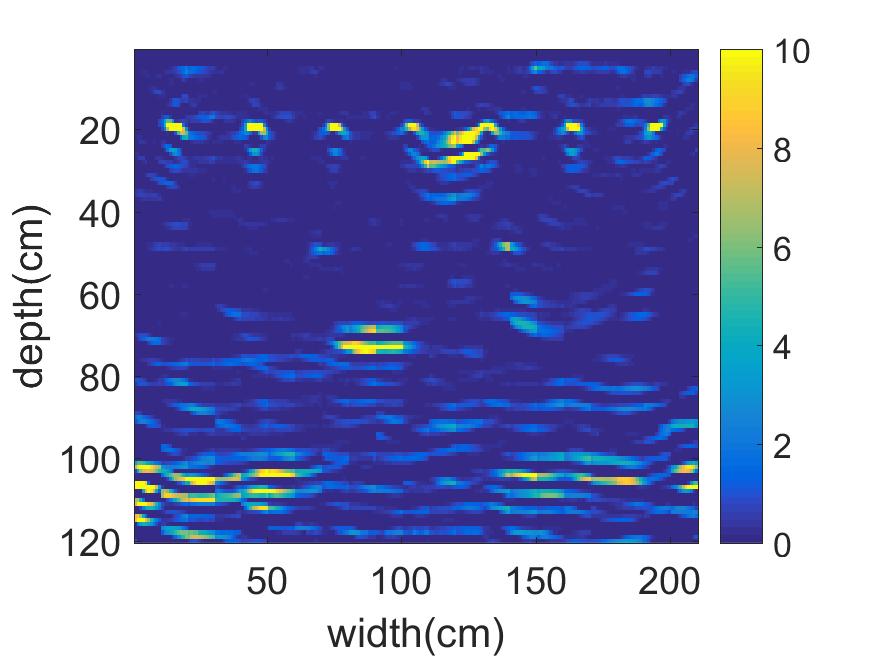}&
			\includegraphics[align = c,width=0.2\textwidth]{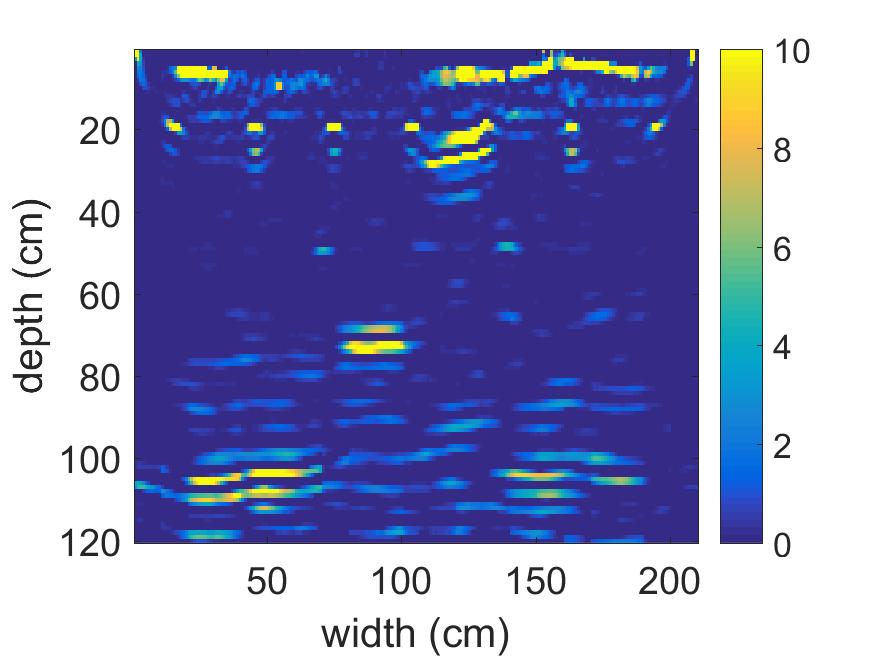} &
			\includegraphics[align = c,width=0.2\textwidth]{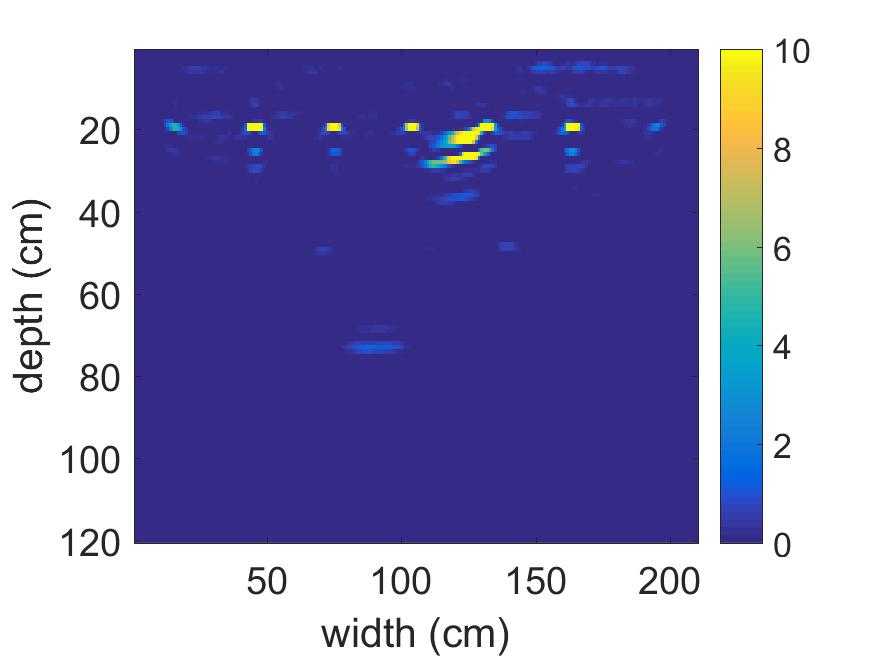}
			
			\tabularnewline
			(e) & (f) & (g) & (h)
			\tabularnewline
			
		\end{tabular}
	\end{center}
	\vspace{-0.15in}
	\caption{\label{MBIRresults}
		\scriptsize
		{A comparison between different settings of MBIR where (a) is the defect diagram of rough-hor-set11, (b) is 2D MBIR reconstruction, (c) is 2.5D MBIR reconstruction with all modifications to the forward and prior models, (d) is 2D MBIR reconstruction without direct arrival signal or shift error estimation, (e) is 2D MBIR reconstruction without shift error estimation, (f) is 2D MBIR reconstruction using regular stitching, (g) is 2D MBIR reconstruction using an isotropic model, (h) is 2D MBIR reconstruction for a constant regularization. The results in (c) shows performance enhancement over the other results.}}
\end{figure*}
\subsection{Results from Modifying the Forward and Prior Models}
\label{modif}
In this section we use the same experimental results from the MIRA data to compare MBIR performance after the modification to the forward and prior models.
The default setting are direct arrival signal elimination, shift error estimation, anisotropic reconstruction, and spatially variant regularization.
The results use the default settings unless otherwise stated.
Fig. \ref{MBIRresults} compares MBIR performance when not using each modification.
Fig. \ref{MBIRresults}b shows 2D MBIR reconstruction.
Fig. \ref{MBIRresults}c shows 2.5D MBIR reconstruction.
Fig. \ref{MBIRresults}d shows 2D MBIR reconstruction without the direct arrival signal modeling.
Fig. \ref{MBIRresults}e shows 2D MBIR reconstruction with the direct arrival signal modeling, but not the shift error estimation.
Fig. \ref{MBIRresults}f shows 2D MBIR reconstruction with an isotropic forward model.
Fig. \ref{MBIRresults}g shows 2D MBIR reconstruction with regular stitching.
Fig. \ref{MBIRresults}h shows 2D MBIR reconstruction with constant regularization.

\subsubsection{Discussion}
Fig. \ref{MBIRresults}c shows the best quality reconstruction with reduced noise and artifacts, and better clarity of the targets.

\section{Conclusion}
\label{conc}
This paper proposed an MBIR algorithm for ultrasonic one-sided NDE. 
The paper showed the derivation of a linear forward model. 
The QGGMRF potential function for the Gibbs distribution prior model was chosen for this problem because it guarantees function convexity, models edges and low contrast regions, and has continuous first and second derivatives.
Furthermore, we proposed modifications to both the forward and prior models that improved reconstruction quality. 
These modifications included direct arrival signal elimination, anisotropic transmit and receive pattern, and spatially variant regularization.
Additionally, a joint-MAP estimate and a 2.5D MBIR were performed to process large multiple scans at once which helps reduce noise and artifacts dramatically compared with results from individual scans.
The research was supported by simulated and extensive experimental results.
The results compared the performance of MBIR with SAFT and $l_1$-norm qualitatively and quantitatively.
The results showed noticeable improvements in MBIR over SAFT and $l_1$-norm in reducing noise and artifacts.

While the results of this paper are promising, it is worth mentioning the need of a non-linear forward model to address the issues due to the complexity of the one-sided UNDE systems, such as reverberation, and acoustic shadowing.

\section{Acknowledgment}
Hani Almansouri and C.A. Bouman were supported by the U.S. Department of Energy. 
S.Venkatakrishnan and Hector Santos-Villalobos were supported by the U.S. Department of Energy’s staff office of the Under Secretary for Science and Energy under the Subsurface Technology and Engineering Research, Development, and Demonstration (SubTER) Crosscut program, and the office of Nuclear Energy under the Light Water Reactor Sustainability (LWRS) program.

\IEEEtriggeratref{8}

\IEEEtriggercmd{\enlargethispage{-0in}}
\bibliographystyle{IEEEtran}
\bibliography{IEEEabrv,IEEEbib}

\end{document}